\documentclass[aps,prb,reprint,superscriptaddress,showpacs,
amsmath,citeautoscript,flushbottom,floatfix]{revtex4-1}

\usepackage{graphicx}
\usepackage{latexsym}
\usepackage{amssymb}

\usepackage{txfonts}\usepackage[full]{textcomp}\usepackage{color}
\newcommand{\vc}[1]{\boldsymbol{#1}}
\newcommand{\GP}{\Gamma^\prime}
\newcommand{\NIO}{\mbox{Na$_2$IrO$_3$}}
\newcommand{\LIO}{\mbox{$\alpha$-Li$_2$IrO$_3$}}
\newcommand{\RuC}{\mbox{$\alpha$-RuCl$_3$}}

\setcitestyle{numbers,square}

\allowdisplaybreaks

\renewcommand{\textfloatsep}{15pt}

\begin{document}

\title{Kitaev-like honeycomb magnets: Global phase behavior and emergent effective models}

\author{Juraj Rusna\v{c}ko}
\affiliation{Central European Institute of Technology,
Masaryk University, Kamenice 753/5, CZ-62500 Brno, Czech Republic}
\affiliation{Department of Condensed Matter Physics, Faculty of Science,
Masaryk University, Kotl\'a\v{r}sk\'a 2, CZ-61137 Brno, Czech Republic}

\author{Dorota Gotfryd}
\affiliation{Institute of Theoretical Physics, Faculty of Physics,
University of Warsaw, Pasteura 5, PL-02093 Warsaw, Poland}
\affiliation{Marian Smoluchowski Institute of Physics,
Jagiellonian University, {\L}ojasiewicza 11, PL-30348 Krak\'ow, Poland}

\author{Ji\v{r}\'{\i} Chaloupka}
\affiliation{Central European Institute of Technology,
Masaryk University, Kamenice 753/5, CZ-62500 Brno, Czech Republic}
\affiliation{Department of Condensed Matter Physics, Faculty of Science,
Masaryk University, Kotl\'a\v{r}sk\'a 2, CZ-61137 Brno, Czech Republic}

\begin{abstract}
Compounds of transition metal ions with strong spin-orbit coupling recently
attracted attention due to the possibility to host frustrated bond-dependent 
anisotropic magnetic interactions.
In general, such interactions lead to complex phase diagrams that may include
exotic phases, e.g. the Kitaev spin liquid.
Here we report on our comprehensive analysis of the global phase diagram of
the extended Kitaev-Heisenberg model relevant to honeycomb lattice compounds
\mbox{Na$_2$IrO$_3$} and \mbox{$\alpha$-RuCl$_3$}.
We have utilized recently developed method based on spin coherent states that
enabled us to resolve arbitrary spin patterns in the cluster ground states
obtained by exact diagonalization.
Global trends in the phase diagram are understood in combination with the
analytical mappings of the Hamiltonian that uncover peculiar links to known
models -- Heisenberg, Ising, Kitaev, or compass models on the honeycomb
lattice -- or reveal entire manifolds of exact fluctuation-free ground states.
Finally, our study can serve as a methodological example that can be applied
to other spin models with complex bond-dependent non-Heisenberg interactions.
\end{abstract}

\date{\today}

\maketitle

\vspace{-2mm}\section{Introduction}\vspace{-2mm}

In contrast to simple examples of Heisenberg magnets discussed in standard
textbooks, frustrated spin systems \cite{Lac11} offer much wider range of
phenomena, including the exotic spin-liquid behavior \cite{Bal10,Sav17} or the
emergence of effective monopoles in spin-ice pyrochlores \cite{Bra01,Cas12}.
The usual sources of frustration are frustrated geometry of the lattice (e.g.
kagom\'{e} \cite{Nor16}) or the presence of longer-range interactions
competing with the nearest-neighbor ones (as e.g. in $J_1$-$J_2$ model
\cite{Cha88,Cha89,Sch17}) and possibly among themselves.
Within the last decade, pseudospin-$\frac12$ systems with frustrated
bond-dependent non-Heisenberg interactions emerging in Mott insulators as a
consequence of spin-orbit coupling (SOC) became a subject of intense research
\cite{Wit14,Nus15,Rau16,Sch16,Tre17,Win17,Her18}. While one of the main
motivations has been a possible realization of the Kitaev honeycomb model
\cite{Kit06}, the presence of additional interactions leads to very rich
magnetic behavior that is particularly attractive as well as challenging to
study. 

The basic element enabling the realization of the above models possessing
bond-dependent anisotropic interactions has been well known for a long time.
It relies on a $d^5$ valence configuration of heavy transition-metal ions with
large SOC which combines the spin $s=\frac12$ and effective orbital angular
momentum $l_\mathrm{eff}=1$ of the hole in the $t_{2g}^5$ configuration into
$J_\mathrm{eff}=\frac12$ Kramers doublet ground state \cite{Abr70,Kha05}. A
direct experimental evidence for the spin-orbital entangled multiplet
structure \cite{Kim08} was obtained e.g. by resonant x-ray scattering on
\mbox{Sr$_2$IrO$_4$} \cite{Kim09} containing $d^5$ Ir$^{4+}$ ions.
It was the seminal theoretical proposal by Jackeli and Khaliullin \cite{Jac09}
that suggested how to exploit the $J_\mathrm{eff}=\frac12$ pseudospins in Mott
insulators with large SOC. Two lines of intense research followed. The first
one focuses on the square lattice case with the resulting Heisenberg
interactions among the pseudospins -- a situation appealingly analogous to
undoped cuprates. Cuprate-like magnetism was indeed found in perovskite
iridates \cite{Kim12} and certain observations support the idea to extend the
analogy to the doped case \cite{Kim14a,Kim16}. Yet bigger excitement was
initiated by a proposal that the honeycomb $J_\mathrm{eff}=\frac12$ compounds
may be close to the Kitaev limit where Ising-like bond-dependent interactions
lead to a spin-liquid ground state. Such an exotic effective spin system may
naturally arise when translating the bond-anisotropic interactions of the $d$
orbitals appearing in Kugel-Khomskii models \cite{Kug82} into the pseudospin
space via the SOC-induced spin-orbital entanglement \cite{Kha05,Che08}.

In the search of materials close to the Kitaev limit, much attention has been
paid to the honeycomb iridates {\NIO}, {\LIO}, and the ruthenate {\RuC}
\cite{Plu14} that is claimed to show signatures of Kitaev physics in the
excitation spectra \cite{Ban16,Ran17,Ban17,Do17}.
However, these compounds were found to host long-range magnetic order instead
-- zigzag type in {\NIO} \cite{Liu11,Ye12,Cho12} and {\RuC}
\cite{Joh15,Sea15,Ban16} and spiral type in {\LIO} \cite{Wil16}. Only very
recently, an evidence for a liquid state was found in a related compound
\mbox{H$_3$LiIr$_2$O$_6$} \cite{Kit18}.
Even though the zigzag phase is present in the phase diagram of the originally
proposed Kitaev-Heisenberg model \cite{Cha13}, later experiments on {\NIO}
showed that it gives an inconsistent ordered moment direction \cite{Chu15} and
additional bond-anisotropic and/or further-neighbor interactions have to be
invoked \cite{Rau14a,Rau14b,Kat14,Yam14,Siz14,Win16}. In the resulting
extended Kitaev-Heisenberg models,
the highly anisotropic interactions lead to complex phase behavior (see
Refs.~\cite{Rau14a,Rau14b,Win17} for examples) or unusual spin
excitation spectra showing e.g. a breakdown of the magnon picture even in the
long-range ordered phase away from the Kitaev limit \cite{Win17b} or
topological features \cite{McC18,Jos18}.

The (extended) Kitaev-Heisenberg models are not limited to the honeycomb
lattice. A large number of other situations have been discussed, including
triangular \cite{Kha05,Bec15,Rou16,Li15,Cat15} and kagom\'{e} \cite{Mor18}
lattice and suitable types of three-dimensional structures such as
experimentally realized hyperhoneycomb \cite{Tak15,Lee15,Kim15,Lee16,Kat16},
harmonic honeycomb \cite{Mod14,Bif14,Kim14c,Lee15,Lee16}, hyperkagom\'{e}
\cite{Oka07,Che08}, fcc \cite{Cao13,Coo15,Acz16}, and pyrochlore lattices
\cite{Kur10,Kim14b}, or hypothetical hyperoctagon lattice \cite{Her14}.
Finally, the concept of Kitaev interactions in pseudospin
$J_\mathrm{eff}=\frac12$ systems has been recently extended to $d^7$ compounds
such as those containing Co$^{2+}$ \cite{Liu18,San18}. 

In general, a thorough inspection of an extended Kitaev-Heisenberg model in
terms of spin structures, excitations etc. through the parameter space is
desired. Apart from theoretical interest, this is mostly in order to establish
it as an effective model for a concrete material and to narrow down the
parameter regime.
Methodologically, the inspection is complicated by the new kind of frustration
stemming from the bond dependence of the interactions. Since exotic features
such as spin-liquid ground states and fractionalized excitations are
``around'', simple approaches -- for instance the Luttinger-Tisza method
\cite{Lut46} or linear spin waves -- often have a limited success and one has
to resort to unbiased numerical methods fully incorporating quantum effects.
Of a great value are also exact symmetry properties, such as dual mappings of
the Hamiltonian utilizing sublattice spin rotations
\cite{Kha05,Cha10,Kim14b,Cha15,Nus15} that proved surprisingly powerful when
establishing and interpreting the phase diagram.

The aim of this paper is to perform a detailed analysis of the phase diagram
of the extended Kitaev-Heisenberg model (EKH) relevant for honeycomb
materials. 
Portions of the phase diagram have been reported before by several studies,
both on the classical level \cite{Rau14a,Siz14,Jan17} as well as including the
quantum effects \cite{Lou15,Shi15,Nis16,Win16,Win17}.
Here we take a global view of the phase diagram, trying to understand its
trends based on the competition/cooperation of the interactions and general
symmetry properties. We also analyze the internal structure of the phases
including the ordered moment direction that is useful when fixing the model
parameters based on experimental data \cite{Chu15,Cha16}.
To this end, we build on previous work \cite{Cha16} and use exact
diagonalization combined with ground-state analysis based on
spin-$\tfrac{1}{2}$ coherent states and complemented by cluster mean field
theory. This allows us to determine the spin structures through the phase
diagram, including the non-collinear ones and estimate the amount of quantum
fluctuations. 
The global analysis revealed two surprising features that underline the
richness of the EKH model and enable a deeper understanding of its phase
behavior:
(i)~sets of exact fluctuation-free ground states forming entire manifolds in
the parameter space,
(ii)~possibility to map part of the phase space of the EKH model to a model
characterized by a single bond-dependent interaction axis. This way several
models of separate interest ``emerge'' from the EKH model: Ising, Kitaev, and
compass \cite{Kug82,Nus15} models as well as their combinations.

The paper is organized as follows: The model and numerical methods are
introduced in Sec.~\ref{sec:model} and Sec.~\ref{sec:methods}, respectively.
Section~\ref{sec:phasediag} contains the phase diagram of the model along with
a discussion of its phases. Section~\ref{sec:T6} analyzes the manifolds of
fluctuation-free ground states. Finally, Section~\ref{sec:ising} is devoted to
the study of the Ising-Kitaev-compass case and its links to EKH model.


\vspace{-2mm}\section{Extended Kitaev-Heisenberg model}\label{sec:model}\vspace{-2mm}

\begin{figure}[b!]
\includegraphics[scale=1.0]{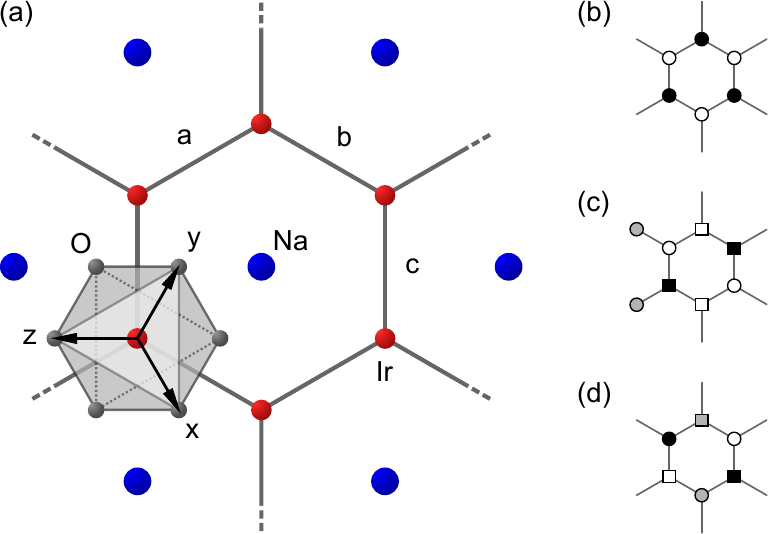}
\caption{
(a)~Honeycomb NaIr$_2$O$_6$ layer. Iridium ions form a honeycomb lattice with
a sodium atom in the middle of each hexagon. Each iridium atom is surrounded
by an octahedron of oxygens; the neighboring octahedra share an edge. The
figure shows also the definition of cubic $x,y,z$ axes and the bond directions
$a,b,c$.
(b-d)~Sublattices of the two-, four- and six-sublattice transformations
$\mathcal{T}_2,\mathcal{T}_4,\mathcal{T}_6$ that reveal the points of hidden
SU(2) symmetry.
}
\label{fig:schem}
\end{figure}

\vspace{-2mm}\subsection{Model Hamiltonian}\vspace{-2mm}

According to the currently available prevailing evidence for honeycomb
materials \cite{Win17} and following Ref.~\cite{Cha16}, we choose to
study the nearest-neighbor extended Kitaev-Heisenberg model
\cite{Kat14,Rau14a,Rau14b} complemented by third-nearest neighbor Heisenberg
exchange.
The nearest-neighbor (NN) part of the model contains -- in addition to the
usual Heisenberg exchange -- all possible anisotropic terms that are allowed
by symmetry of the trigonally distorted honeycomb lattice \cite{Kat14,Cha15}.
It is most conveniently expressed in cubic coordinates $x$, $y$, $z$
introduced in Fig.~\ref{fig:schem}(a) that allow to easily incorporate the
discrete rotational $C_3$ symmetry. For a bond of $c$ direction, the
Hamiltonian contribution reads as
\begin{multline}\label{eq:Hcubicc}
  \mathcal{H}_{ij}^{(c)} =
  J\, {\vc S}_i \cdot {\vc S}_j + K\, S_i^z S_j^z \\
  \!+\!\Gamma(S_i^x S_j^y \!+\! S_i^y S_j^x)
  \!+\!\Gamma'(S_i^x S_j^z \!+\! S_i^z S_j^x \!+\! S_i^y S_j^z \!+\! S_i^z S_j^y)
\end{multline}
whereas the contributions for the other bond directions are obtained by a
cyclic permutation of the spin components $S^x,S^y,S^z$. The $J$ and $K$
terms alone constitute the Kitaev-Heisenberg model \cite{Jac09,Cha10} that has
been subject to extensive studies
\cite{Cha10,Cha13,Got17,Pri12,Sch12,Reu11,Ire14,Goh17,Tro13,Tro14} 
and still serves as a prototype model to capture a departure from the 
Kitaev physics. 
In light of experimental data \cite{Chu15}, it has been generally recognized
that further anisotropic terms are needed, leading to the addition of the
$\Gamma$ and $\GP$ terms introduced in Refs.~\cite{Rau14a,Rau14b,Kat14}.
When studying the phase diagram we keep signs of $J$ and $K$ flexible and fix
the signs $\Gamma>0$ and $\Gamma'<0$ following the ab-initio calculations as
well as the perturbative evaluation of the effective interactions
\cite{Win16,Rau14b}. According to the latter one, small negative $\Gamma'$ should
correspond to a trigonal compression of the lattice \cite{Rau14b}, observed in
Na$_2$IrO$_3$ \cite{Cho12,Ye12}.
Moreover, several ab-initio studies have evaluated the importance of
further-neighbor couplings (see e.g. \cite{Yam14,Win16}). In
Ref.~\cite{Win16}, the effective spin Hamiltonians for {\NIO}, {\RuC},
and {\LIO} were constructed using a combination of DFT and cluster exact
diagonalization that equally treated interactions up to third nearest
neighbors. Among the further-neighbor interactions, a significant value of
$J_3>0$ was found for all three compounds, which leads us to the complete
model considered here
\begin{equation}\label{eq:Hfull}
  \mathcal{H} = 
  \sum_{\langle ij \rangle \in \mathrm{NN}} \mathcal{H}_{ij}^{(\gamma)}
  +\!\!\!\sum_{\langle ij \rangle \in 3^{\mathrm{rd}}\,\mathrm{NN}}
  \!\!\!\!\!\!J_3 \vc S_i\cdot \vc S_j \;.
\end{equation}

\vspace{-2mm}\subsection{Hidden symmetries of the model}\vspace{-2mm}

The NN part of the above model ($J_3=0$) has rich symmetry properties explored
in detail in the previous work \cite{Cha15}. 
First, it supports a self-dual transformation $\mathcal{T}_1$ that corresponds
to a global $\pi$-rotation of the spins around the axis perpendicular to the
honeycomb plane. Such a transformation fully preserves the form of the
Hamiltonian but replaces the values of the parameters $JK\Gamma\Gamma'$ by
another set of values.  
Second, Ref.~\cite{Cha15} has also identified a number of special parameter
combinations -- the points of ``hidden'' SU(2) symmetry in the parameter space
-- for which the NN model maps to ferromagnetic (FM) or antiferromagnetic (AF)
Heisenberg model on the honeycomb lattice.  This is achieved by employing
either \mbox{two-,} \mbox{four-,} or six-sublattice coverings of the honeycomb
lattice as depicted in Fig.~\ref{fig:schem}(b)-(d) and performing selected
sublattice-dependent rotations of the spins. The neighboring spins that belong
to different sublattices are therefore rotated in a different fashion and the
interaction among those spins takes a modified form, in certain cases the
simple Heisenberg one. For these particular cases, the seemingly anisotropic
model is thus exactly equivalent to the Heisenberg model on the honeycomb
lattice. By using the same transformation backwards, we can exploit the known
properties of Heisenberg model obtaining thereby e.g. the ordering pattern or
excitation spectra at the points of ``hidden'' SU(2) symmetry. Due to the
sublattice structure of the transformation, the simple ordering patterns of
Heisenberg FM or AF transform to more complex ones such as stripy, zigzag or
even non-collinear vortex pattern.

As a well-known example, we can consider the Kitaev-Heisenberg model with the
parameters satisfying the relation $K=-2J$ and the four-sublattice covering
shown in Fig.~\ref{fig:schem}(c). Keeping the spins at the sites marked by
$\square$ unrotated, and applying $\pi$-rotations around the $x$, $y$, or $z$
axes to the spins at the sites attached to the $\square$ sites by the $a$,
$b$, or $c$ bond, respectively, we obtain the Heisenberg Hamiltonian
$\mathcal{H}=-J\sum_{\langle ij\rangle}{\vc S'}_i\cdot{\vc S'}_j$ in the
rotated spin variables $\vc S'$.  In the notation of Ref.~\cite{Cha15}, this
transformation is called $\mathcal{T}_4$. The other possibilities include
two-sublattice transformation $\mathcal{T}_2$, the six-sublattice
$\mathcal{T}_6$, and the combinations $\mathcal{T}_1\mathcal{T}_4$ and
$\mathcal{T}_2\mathcal{T}_6$. All these points of ``hidden'' SU(2) symmetry
summarized in Table~I and Fig.~3 of Ref.~\cite{Cha15} provide exact
reference points in the parameter space and will be extensively utilized in
the present study.


\vspace{-2mm}\section{Methods}\label{sec:methods}\vspace{-2mm}

To solve the model, we use the standard Lanczos exact diagonalization (ED)
technique employing a finite cluster \cite{Ave13}. The calculated cluster
ground state is subsequently analyzed utilizing spin-$\frac12$ coherent states
\cite{Cha16} as detailed below. The ED technique is complemented by the
cluster mean-field theory (CMFT). This combination is useful for a global
characterization of the phase diagram -- ED gives the ground state
characteristics such as energies and spin correlations, the analysis based on
spin-$\frac12$ coherent states enables to better assess the ordering patterns
and the direction of magnetic moments, and CMFT supplements this information
by the length of the ordered moments which is not directly accessible by ED.

In both cases we use a hexagonal 24-site cluster with periodic boundary
conditions applied. This cluster has a fully symmetric shape and supports all
the phases with hidden SU(2) symmetry \cite{Cha15}. It is therefore expected
to provide a fair environment for the competition of the phases, with the
exception of the possible spiral phases that are forced to fit the periodic
boundary conditions and may be thus slightly suppressed. 
In this specific case we have extended our ED analysis to 32-site clusters.

\vspace{-2mm}\subsection{Spin-$\frac{1}{2}$ coherent states for non-collinear phases}\label{sec:methcoh}\vspace{-2mm}

The analysis of the exact ground state of the cluster obtained by ED presents
a challenge -- the cluster ground state does not spontaneously break symmetry
but instead contains a linear combination of all the degenerate spin
configurations. To resolve the dominant configuration and obtain the direction
of the pseudospins from the ED ground state, we follow Ref.~\cite{Cha16}
and employ spin-$\frac{1}{2}$ coherent states. Such a state, polarized in a
direction given by spherical angles $\theta,\phi$ is given by:
\begin{equation}
  |\theta,\phi\rangle=e^{-i\phi S^z}e^{-i\theta S^y}|\!\uparrow\,\rangle, 
\end{equation}
where we make a standard choice of cubic $z$ direction as the quantization
axis. The cluster spin-coherent state is then a direct product of coherent
states on each site $j$:
\begin{equation}\label{eq:cohstate}
  |\Psi\rangle = \prod_{j=1}^{N}|\theta_j,\phi_j\rangle.
\end{equation}
This state can be understood as a classical (fluctuation-free) spin pattern
with the individual spins pointing in the directions determined by the angles
$\theta_j,\phi_j$. By calculating the overlap $\langle\Psi|\mathrm{GS}\rangle$
and maximizing its absolute value by varying the angles, we can identify the
classical pattern that best fits the exact ground state $|\mathrm{GS}\rangle$.

For collinear phases (in the case of EKH model, these are FM, AF, zigzag and
stripy) the cluster spin-coherent state is captured by a single pair
$(\theta,\phi)$, which makes it easy to find the moment direction by
inspecting the probability map
$P(\theta,\phi)=|\langle\Psi|\mathrm{GS}\rangle|^2$ and finding the maximum.
However, already the analysis of hidden SU(2) points revealed the existence of
several non-collinear ``vortex'' phases in the phase diagram of the EKH model
\cite{Cha15}. In the general case, the probability
$P=|\langle\Psi|\mathrm{GS}\rangle|^2$ has to be maximized with respect to all
$2N$ angles. For our cluster with $N=24$ sites, this poses a nontrivial
computational problem of global optimization in 48-dimensional space. To this
end, we use the particle swarm method for global optimization, which yields a
result further refined by a local optimization algorithm.

The demanding task can be partly avoided by estimating in advance the
parameter windows where non-collinear phases can be found. This can be
achieved by first finding the optimal spin configuration among the collinear
ones and calculating the full Hessian matrix of second derivatives (with
respect to all 48 angular parameters) for such a configuration. The potential
instability of the collinear phase can be identified by analyzing the
eigenvalues of this Hessian matrix.


\vspace{-2mm}\subsection{Cluster mean-field theory}\label{sec:methCMFT}\vspace{-2mm}

Similarly to ED, within CMFT we periodically cover the lattice by copies of a
given cluster. The bonds connecting the cluster copies (external bonds) are
treated in a mean-field approximation, replacing the contributions to the bond
Hamiltonian according to the recipe
\begin{equation}
  S^\alpha_i S^\beta_j\approx\langle S^\alpha_i\rangle S^\beta_j + 
  S^\alpha_i \langle S^\beta_j\rangle - 
  \langle S^\alpha_i\rangle \langle S^\beta_j\rangle,
\end{equation}
while the internal bonds of the cluster are kept fully quantum \cite{Got17}.
The mean-field approximation generates effective magnetic fields acting at the
outer sites of the cluster and polarizing the cluster ground state to be
determined by ED. The polarizing fields depend on the averages $\langle
S^\alpha_i\rangle$ measured on the polarized ground state which leads to a
selfconsistent problem with much higher computational demands than the pure
ED. On the other hand, by explicitly breaking the ground state symmetry, the
CMFT method allows to directly determine the ordering pattern and estimate the
ordered moment length.

The introduction of the mean-field boundary makes the sites of the cluster
nonequivalent. In combination with the highly anisotropic bond-dependent
interactions, the spin structures show a tendency towards various forms of
artificial canting. To prevent this, we limit ourselves to the case of
collinear spin structures and follow the approach described in
Ref.~\cite{Got17}, where an averaged ordered moment through the cluster
is taken and distributed on the boundary sites following a particular ordering
pattern.



\vspace{-2mm}\section{Global phase diagram}\label{sec:phasediag}\vspace{-2mm}

\begin{figure}[t]
\includegraphics[scale=1.046]{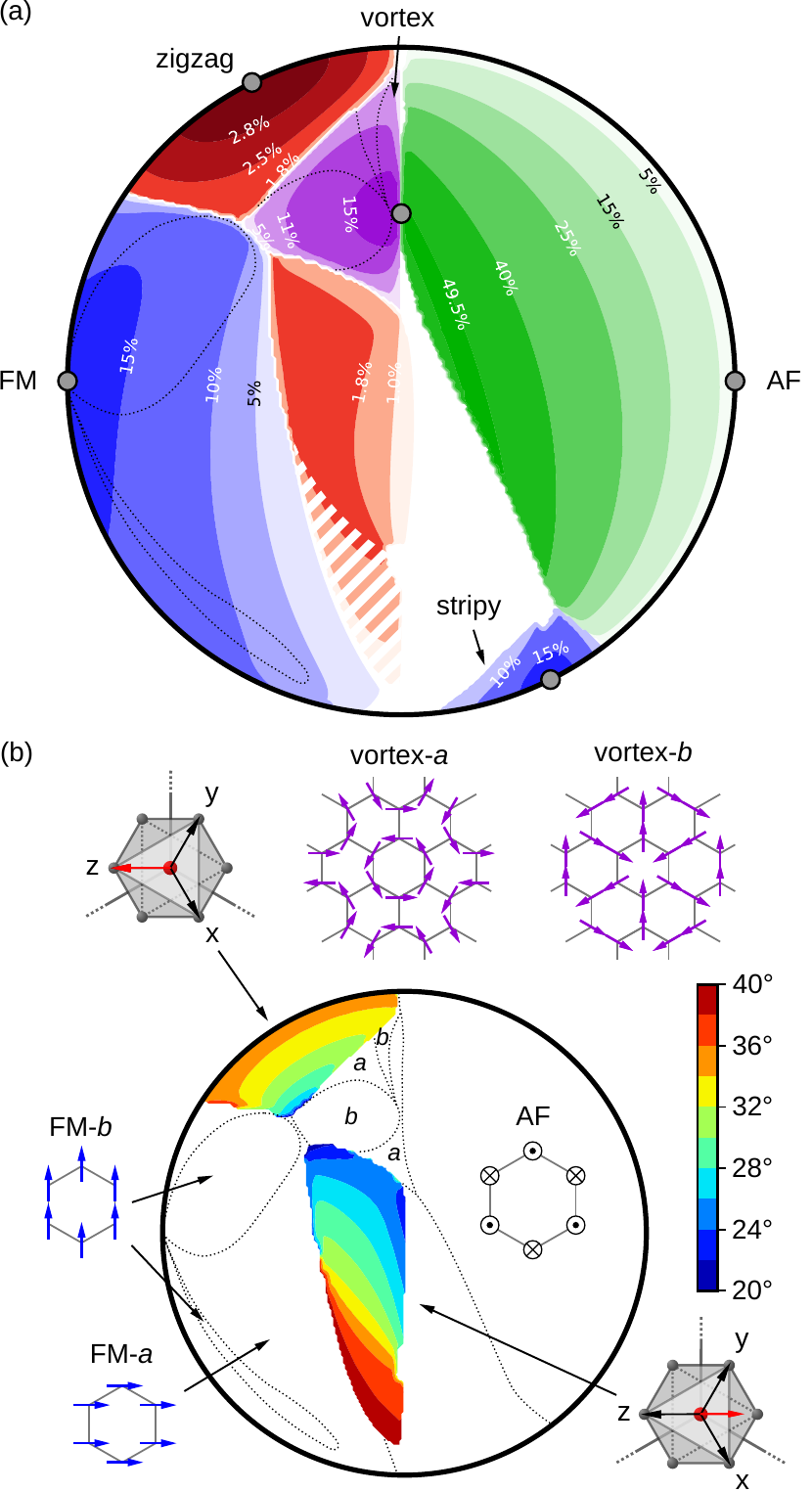}
\caption{
(a)~Phase diagram of the extended Kitaev-Heisenberg model with
$\Gamma^\prime=J_3=0$ using the parametrization $J=\cos\varphi\sin\theta$,
$K=\sin\varphi\sin\theta$, $\Gamma=\cos\theta$ with $\varphi\in[0,2\pi]$ being
the azimuthal angle and $\theta\in[0,\pi/2]$ the radial coordinate measured as
a distance from the center of the circle. Color intensity and contours show
the probabilities of classical spin patterns for the respective phases. Dashed
lines separate distinct regions within one phase. White areas represent
regions where no clear signatures of a long-range ordered phase were detected
using the 24-site cluster. The gray dots indicate points of (hidden) SU(2)
symmetry. The hatched part of the central region with a large probability of
the zigzag pattern shows a tendency to form a non-collinear spin arrangement.
(b)~The angle of the ordered moments to the honeycomb plane for the zigzag
phase -- in the upper region, the moment points near the cubic $z$ direction
(assuming zigzags running along $a$ and $b$ bonds), whereas in the central
region it is located between $x$ and $y$ axes. The panel also depicts the
in-plane spin patterns for distinct regions (labeled as \textit{a} and
\textit{b}) within vortex and FM phases and the out-of-plane pattern of the AF
phase. Further details can be found in Fig.~\ref{fig:parevol}(e) of
Appendix~\ref{sec:appendixPD}.
}
\label{fig:prob}
\end{figure}

By optimizing the spin configurations using the methods described in the
previous section and evaluating the corresponding probabilities, we are able
to construct a detailed phase diagram of the model. We present the slices
through the phase diagram using a common parametrization for the main
interactions \cite{Rau14a}, that is $J=\cos\varphi\sin\theta$,
$K=\sin\varphi\sin\theta$, $\Gamma=\cos\theta$ with $\varphi\in[0,2\pi]$ and
$\theta\in[0,\pi/2]$. This way all the $J$, $K$ sign combinations and
interaction strength ratios for positive $\Gamma\geq 0$ are explored. The
remaining model parameters $\Gamma^\prime$ and $J_3$ are kept fixed for a
given slice. Fig.~\ref{fig:prob}(a) shows the phase diagram for
$\Gamma^\prime=J_3=0$, which is the special case of the $JK\Gamma$ model,
first analyzed in Ref.~\cite{Rau14a}. We shall now use this diagram to
survey the main properties of the phases and move on to their evolution with
$\GP$ and $J_3$ afterwards. The reader may also consult 
Appendix~\ref{sec:appendixPD} containing an extensive set of phase diagram
slices for selected $\GP$ values.

\vspace{-2mm}\subsection{Collinear phases of the $JK\Gamma$ model}\label{sec:collin}\vspace{-2mm}

We first focus on the simpler collinear phases which occupy most of the phase
diagram. Two phases, FM and AF, dominating Fig.~\ref{fig:prob}(a) are directly
linked to Heisenberg points. Though they may seem trivial at the first sight,
our inspection revealed their interesting internal structure due to the
complex interplay of the bond-anisotropic interactions. We start with the FM
phase, which is expected to be most accessible due to the small amount of
quantum fluctuations. In the $JK$ limit (the outer circle of the diagram), the
ordered moments point along one of the three cubic directions $x,y,z$ selected
by virtue of the ``order-from-disorder'' mechanism on top of isotropic
classical energy \cite{Tsv95,Cha16}. Since the cluster ground state is a
superposition of the six degenerate possibilities, the probability approaching
the value $\frac16$ near the FM point indicates vanishing quantum
fluctuations. With the presence of the $\Gamma$ term, the magnetic moment is
quickly pushed into the honeycomb plane, lying either directly within the
plane or close to it (with $\lesssim 10^\circ$ deviation, see also
Fig.~\ref{fig:parevol}(e) in Appendix~\ref{sec:appendixPD}). This can be
understood by evaluating the classical energy for the FM phase:
$E_\mathrm{class}\propto 3J+K-\Gamma+\Gamma(n_x+n_y+n_z)^2$, where the unit
vector $\vc n=(n_x,n_y,n_z)$ represents the moment direction. The honeycomb
plane is thus preferred by $\Gamma$ interaction on a classical level which
makes it easy to outweigh the fluctuation-selected cubic axis. A small
$\Gamma$ value of the order $10^{-2}$ to $10^{-1}$ of the dominant $JK$ is
typically sufficient to achieve this with the value dropping even lower near
the FM Heisenberg point. Within the honeycomb plane, moments point either in
the bond direction, or perpendicular to the bond in two separate regions of
the FM phase [see Fig.~\ref{fig:prob}(b)]. In accord with the intuition,
departing from the FM Heisenberg point, quantum fluctuations intensify,
lowering thus the plotted probability. 

Linked to the FM phase by means of the four-sublattice transformation
($\mathcal{T}_4$ in the notation of Ref.~\cite{Cha15}) is the stripy
phase. Its hidden FM nature is manifested by a large probability, reaching
$\frac16$ at the hidden SU(2) point $K=-2J<0$ that is an image of the FM
Heisenberg point in the $\mathcal{T}_4$ mapping. In contrast to the FM phase,
the magnetic moment direction is tied to the vicinity of the cubic axes
throughout the stripy phase, lifting a bit with increasing $\Gamma$ instead of
moving to the honeycomb plane. This is because the $\Gamma$ interaction is not
compatible with the $\mathcal{T}_4$ transformation and acts differently here.

In the AF phase, the moment direction is classically degenerate in the $JK$
limit, and the cubic directions are chosen again by the
``order-from-disorder'' mechanism. The addition of the $\Gamma$ anisotropy
fixes now the moments in the $(111)$ direction -- perpendicular to the
honeycomb plane. This state minimizes the classical energy including $\Gamma$
contribution: $E_\mathrm{class}\propto -3J-K+\Gamma-\Gamma(n_x+n_y+n_z)^2$.
Similarly to the FM phase, the fluctuation energy selecting the cubic
directions is small and the change to the $(111)$ direction occurs already at
a minute $\Gamma$ of the order $10^{-4}$ to $10^{-2}$ of the dominant $JK$
with the critical value of $\Gamma$ decreasing to zero at the AF Heisenberg
point.  Going deeper into the AF phase, the probability of the classical
N\'{e}el configuration increases with $\Gamma$ steadily, peaking at $\frac12$
on a line near the circle center that starts at the $K=\Gamma$ hidden SU(2)
symmetry point. For the (111) AF state, there are two equivalent
configurations of the moments, meaning that the peaking probability of 50\%
represents a classical state without any quantum fluctuations. Indeed, as we
later explicitly demonstrate in Sec.~\ref{sec:T6}, terms that would lead to
quantum fluctuations are present but their remarkable cancellation for the
particular order causes the highly anisotropic model to support a
fluctuation-free AF state on an entire manifold of its parameter space. The
same AF phase may thus be represented by fluctuation-free ground states as
well as those with significant quantum fluctuations, depending on the location
in the parameter space.

Analogous to the FM/stripy case, $\mathcal{T}_4$ maps the AF Heisenberg point
to the hidden SU(2) point \mbox{$K=-2J>0$}. The top zigzag region of the phase
diagram extends around this point; in the $JK$ limit, the moment direction
coincides again with one of the cubic directions. Adding further anisotropy
with increasing $\Gamma$, the moments are pushed continuously towards the
honeycomb plane, as shown in Fig.~\ref{fig:prob}(b). 

Of a greater experimental relevance is the second zigzag phase near the center
of the phase diagram. It is also linked to a hidden SU(2) point which,
however, occurs at finite $\Gamma'<0$ \cite{Cha15}. In this phase, the moment
direction is located roughly between the cubic $x$ and $y$ axes, near the
direction found experimentally \cite{Chu15}. We will show later, that it is
this zigzag region that largely expands and dominates the phase diagram after
the inclusion of $\GP$ and/or $J_3$ coupling terms. A comprehensive discussion
of the moment direction in both zigzag phases in the context of the
experimental data can be found in Ref.~\cite{Cha16}.


\vspace{-2mm}\subsection{Vortex phase}\vspace{-2mm}

The vortex phase is a non-collinear phase ``emanating'' from the most peculiar
hidden SU(2) symmetry point of the model that is revealed by a six-sublattice
spin rotation $\mathcal{T}_6$ of Ref.~\cite{Cha15}. $\mathcal{T}_6$ maps
the ferromagnetic $J<0$ Heisenberg model to EKH model at the parameter point
$J=0$, $K=\Gamma>0$ indicated in Fig.~\ref{fig:prob}(a). Owing to its hidden
FM nature and six degenerate spin configurations, the optimized probabilities
reach $\tfrac{1}{6}$ in the vicinity of this exact vortex point, and
continuously decrease with the departure away from it. 

The phase comprises regions with two different most probable classical
configurations of moments labeled as \mbox{vortex-\textit{a}} and
\mbox{vortex-\textit{b}} in Fig.~\ref{fig:prob}(b). Let us note, however, that
these two patterns have very close probabilities and are continuously
connected, implying a presence of a soft mode oscillating between them.
In partial agreement with the classical treatment \cite{Rau14a}, spins are
found to lie within or close to the honeycomb plane. The
\mbox{vortex-\textit{b}} pattern is always planar while in the
\mbox{vortex-\textit{a}} regions near the boundary with AF or zigzag phase,
the spins start to tilt away from the honeycomb plane in a staggered AF
fashion. The tilt is largest in the right part of the vortex phase [see
Fig.~\ref{fig:parevol}(e)] which we interpret as the proximity effect of the
robust AF order with the moments perpendicular to the honeycomb plane.

A deeper understanding of the internal structure of the vortex phase is
possible by utilizing four reference points where the EKH maps to simpler
models. 
One of them is the vortex SU(2) point in $\Gamma'=0$ slice. The freedom
associated with the selection of the ordered moment direction in the hidden FM
at this point creates a continuous family of degenerate patterns including
\mbox{vortex-\textit{a}} and \mbox{vortex-\textit{b}}. 
Another hidden SU(2) symmetry point but of AF nature is found for
$\Gamma'\approx-0.5$ at the opposite edge of the vortex phase [see
Fig.~\ref{fig:parevol}(j) of Appendix~\ref{sec:appendixPD}]. It is associated
with $\mathcal{T}_2\mathcal{T}_6$ transformation of Ref.~\cite{Cha15}.
For a planar structure, the staggering of the hidden AF order is compensated
by the two-sublattice $\pi$-rotation $\mathcal{T}_2$ such that this point
supports the same \mbox{vortex-\textit{a}} and \mbox{vortex-\textit{b}}
patterns as for the $\Gamma'=0$ hidden FM point associated with just
$\mathcal{T}_6$. However, in contrast to the latter point, the corresponding
state has pronounced quantum fluctuations because of the hidden AF nature.
Departing away from the hidden FM point or the hidden AF point, the degeneracy
is lifted and one of the configurations is chosen as the energetically most
favorable. Here the proximity to the remaining two reference points decides.
As we find in Sec.~\ref{sec:ising}, the point $K=\Gamma=-J>0$ (the ``meeting''
point of four phases) corresponds to a FM compass model on the honeycomb
lattice while at another nearby point with $K>0$, $J=\Gamma>0$, and
$\Gamma'<0$, the model maps to AF compass-like model with the interaction
direction perpendicular to the bond. These two \mbox{compass(-like)} models 
prefer patterns \mbox{vortex-\textit{b}} and \mbox{vortex-\textit{a}}, 
respectively, which qualitatively explains the location of 
\mbox{vortex-\textit{a,b}} subphases. 


\vspace{-2mm}\subsection{Remaining phases of the $JK\Gamma$ model}\label{sec:white}\vspace{-2mm}

The remaining parts of the phase diagram slice for $\Gamma'=0$ and $J_3=0$
[kept white in Fig.~\ref{fig:prob}(a)] are to a small extent occupied by the
two known Kitaev spin liquids associated with the FM and AF Kitaev points.
Here the optimization of spin-$\frac12$ coherent states described in
Sec.~\ref{sec:methcoh} finds a large number of configurations consisting of
aligned/contra-aligned pairs of the nearest-neighbor spins, as appearing in
classical $S\rightarrow \infty$ limit of the Kitaev model \cite{Bas08,Rou18}
(see also Appendix \ref{sec:appendixSL} for several details concerning the
behavior of the method in the presence of Kitaev spin liquids).

\begin{figure}[t!]
\includegraphics[scale=1.0]{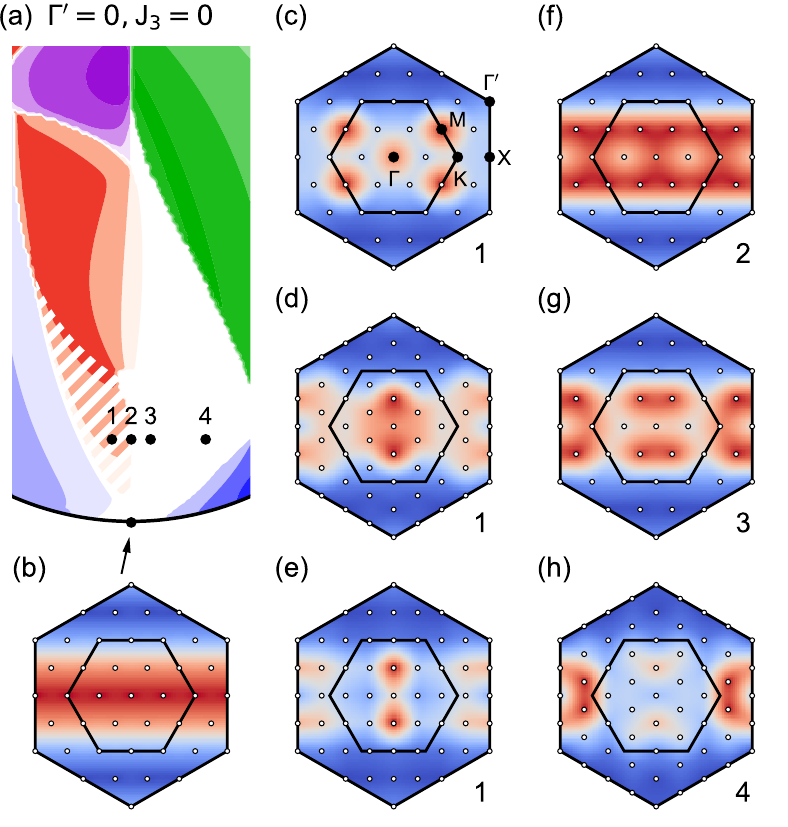}
\caption{
(a)~Position of the four selected points $1-4$ in the phase diagram.
In addition, the FM Kitaev point is taken as a reference.
(b)--(h)~$\langle S_{-\vc{q}}^zS^z_{\vc{q}}\rangle$ correlations at the
selected parameter points calculated for the 24-site cluster (b,c,f,g) and 32-site
clusters of hexagonal (d,h) and rectangular shape (e). The nearest-neighbor correlations
in the liquid state are manifested by a wave-like pattern [panel (b)] -- such
a pattern seems to be present as a ``background'' in the other maps (c--h) as
well. 
At point $1$, the larger 32-site clusters already support incommensurate correlations
[panels (d),(e)], while the 24-site cluster shows zigzag-like correlations
[panel (c)] though collinear zigzag is not the most probable configuration
anymore. Incommensurate correlations are visible at the 24-site cluster for point
$3$ [panel (g)] and merge with the zigzag ones on the $K-\Gamma$ line [panel
(f)]. Deeper in the white region, the incommensurate wavevector moves out of
the first Brillouin zone [panel (h)].
All the panels (b--h) show the available $\vc{q}$-resolution for the 
given cluster. In panel (c), the high-symmetry points in the Brillouin zone
are labeled.
}
\label{fig:zigzagspiral}
\end{figure}

However, much bigger portion of the phase diagram is taken by the white region
in the lower central part which shows a particularly puzzling behavior. Parts
of it were suggested earlier to host incommensurate phases
\cite{Rau14a,Shi15}. The vertical $J=0$ line seems to play a special role as
it clearly separates the middle zigzag as well as the vortex region from the
other phases on the right [see Fig.~\ref{fig:prob}(a) and the detail in
Fig.~\ref{fig:zigzagspiral}(a)]. The $K-\Gamma$ model corresponding to the
$J=0$ line has been recently studied separately and its ground state for
ferromagnetic $K$ was found to bear signatures of a spin liquid
\cite{Goh18,Sta18}.

Using the method of Sec.~\ref{sec:methcoh} for the above region, we find
tendencies to form complex spin structures, though the probability of such
configurations is quite small, hinting towards a possibility of phase(s)
without a long-range order.
Interestingly, the region with a large probability of the collinear zigzag
structure is also partially unstable towards a formation of a non-collinear
spin arrangement -- see the hatched pattern in Fig.~\ref{fig:prob}(a) or
Fig.~\ref{fig:zigzagspiral}(a).
Although the clusters accessible to ED are not in general large enough to
properly capture potential spin orderings with large unit cells, we still try
to provide a further analysis based on momentum-space correlations. Here we
utilize two more clusters in ED, a 32-site cluster of a hexagonal shape and a
rectangular one ($4\sqrt{3}\times 6$ in lattice spacings), in addition to our
default 24-site cluster.

Fig.~\ref{fig:zigzagspiral}(a) shows the positions of four parameter points
selected for a comparison: $1$ in the unstable zigzag region, $2$ on the
$K-\Gamma$ boundary, $3$ in the expected spiral phase close to $J=0$ line, and
point $4$ deeper in the expected spiral phase. The FM Kitaev point is added
for reference. 
Plotted in Fig.~\ref{fig:zigzagspiral}(b)--(h) are the maps of the equal-time
spin-spin correlation function $\langle S_{-\vc{q}}^zS^z_{\vc{q}}\rangle$.
It should be emphasized, that the cluster ground states do not spontaneously
break symmetry and contain e.g. a linear combination of several ordering
patterns that differ by the direction of the ordering wavevector and hence the
ordered moment direction. The selection of the spin component of the
correlation function then provides access to various components of this
combination. For the hexagonal clusters, where a rotation by $2\pi/3$ is in
effect just a cyclic permutation among the $S^x$, $S^y$, and $S^z$ components,
the other correlation functions $\langle S_{-\vc{q}}^xS^x_{\vc{q}}\rangle$ and
$\langle S_{-\vc{q}}^yS^y_{\vc{q}}\rangle$ are merely $2\pi/3$-rotated copies
of the maps shown in Fig.~\ref{fig:zigzagspiral}. 

By combining various sets of maps from Fig.~\ref{fig:zigzagspiral}, several
trends can be illustrated:\\
(i)~The wave-like background identical to momentum-represented
nearest-neighbor correlations of the Kitaev liquid
[Fig.~\ref{fig:zigzagspiral}(b)] is universally present at all points, less
apparently in the case of peaked structures on top of the background because
of an extended colorscale range.\\
(ii)~Panels (c), (d), and (e) show the influence of the cluster size and shape
at the parameter point $1$ that we demonstrate now to be in the incommensurate 
region. For the smallest 24-site cluster, the correlation map in
Fig.~\ref{fig:zigzagspiral}(c) still includes peaks located at the $M$ momenta
which corresponds to a zigzag arrangement. However, using the method of
Sec.~\ref{sec:methcoh}, the zigzag pattern is found unstable which already
hints towards another type of ordering. This is fully revealed by the larger
32-site clusters. By providing a denser momentum-space coverage, they enable
the preferred incommensurate state to develop
[Figs.~\ref{fig:zigzagspiral}(d),(e)]. 
The difference between Fig.~\ref{fig:zigzagspiral}(d) and
Fig.~\ref{fig:zigzagspiral}(e) is an effect of the cluster shape.  The
symmetric hexagonal 32-site cluster [panel (d)] supports 
three degenerate directions for
the ordering wavevector that coexist in the ground state (two of them
visible aside the main maxima near the
Brillouin zone center), while the rectangular shape of the second
32-site cluster selects only one of those directions [panel (e)].\\
(iii) Panels (c), (f), and (g) demonstrate, for the 24-site cluster, the
evolution from commensurate correlations [point~1,
Fig.~\ref{fig:zigzagspiral}(c)] to incommensurate ones [point~3,
Fig.~\ref{fig:zigzagspiral}(g)] found in the white region. At the boundary
point $2$ with $J=0$, the corresponding states show a level crossing and we
obtain the average spin-correlation pattern displayed in Fig.~\ref{fig:zigzagspiral}(f)
resembling that of the Kitaev point.\\
(iv) Panels (d) and (h) illustrate, for the symmetric 32-site cluster, the
transfer of the incommensurate wavevector from the inside of the first
Brillouin zone [point $1$, Fig.~\ref{fig:zigzagspiral}(d)] to the outside 
[point $4$, Fig.~\ref{fig:zigzagspiral}(h)] when moving in the direction of
positive $J$. This trend was also obtained by classical Monte-Carlo
simulations \cite{Rau14b}.

In conclusion, the studied region of the phase diagram shows a complex
behavior with the spin-correlations indicating tendencies towards various
incommensurate orders. However, the common wave-like background to the
spin-correlations suggests a presence of strong liquid-like features.


\vspace{-2mm}\subsection{Effect of nonzero $\Gamma^\prime$ and $J_3$ parameters}\vspace{-2mm}

\begin{figure}[t]
\includegraphics[scale=0.96]{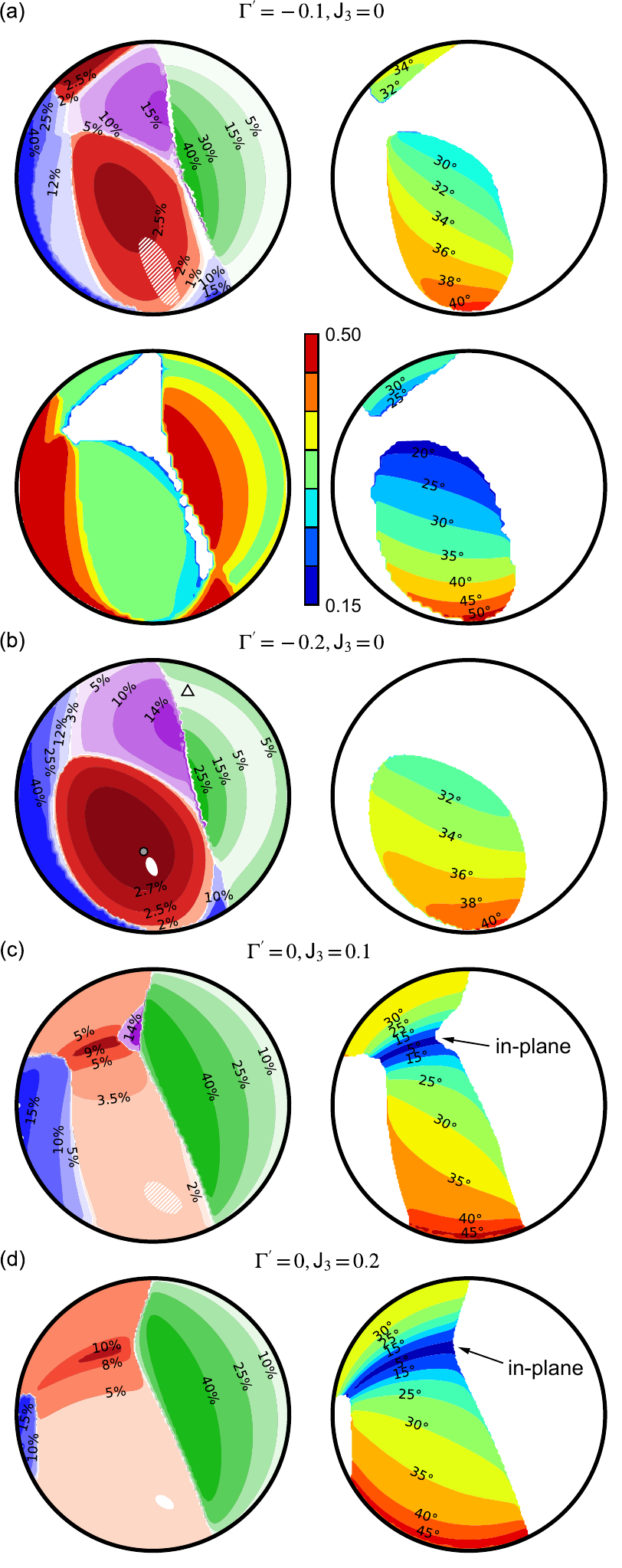}
\caption{
Phase diagrams for nonzero values of $\Gamma'$ and $J_3$ represented by
probabilities of optimized collinear and vortex spin patterns in the ED ground
state (left) and, focusing on zigzag phases, by the angle of the moments to
the honeycomb plane (right). Hatched/white areas in the zigzag phase indicate
the instability of the collinear pattern. Bottom part of panel (a) also shows
the ordered moment length calculated by CMFT (left) and the angle to the
honeycomb plane in zigzag phases (right). Shown in panel (b) are projected
positions of a hidden SU(2) point (gray \textbullet) and a compass-like point
($\vartriangle$).
}
\label{fig:angles}
\end{figure}

We shall now investigate the evolution of the phases found in the
$\Gamma^\prime=J_3=0$ slice of the phase diagram when the parameters
$\Gamma^\prime$ and $J_3$ are varied. As argued in Sec.~\ref{sec:model}, we
limit ourselves to the experimentally most relevant case of small $\Gamma'<0$
and $J_3>0$. Additional data to establish a fuller picture are presented in
Appendix~\ref{sec:appendixPD}. The observed trends can be successfully
explained either simply by considering the classical energy or, more
fundamentally, correlated with the positions of the points of special symmetry
in the parameter space, as inspected in Ref.~\cite{Cha15} (points of
hidden SU(2) symmetry) and the following sections \ref{sec:T6} and
\ref{sec:ising}.

Fig.~\ref{fig:angles}(a),(b) shows phase diagrams for two moderate values of
$\Gamma^\prime<0$. Most notable effect of negative $\Gamma'$ is the large
expansion of the vortex phase and mainly of the central zigzag phase. 
The former trend can be understood as a proximity effect of the point where
the EKH model maps to AF compass-like model (to be analyzed in
Sec.~\ref{sec:ising}). Its position is at $\Gamma'\approx -0.6$ in the chosen
parametrization and the projection to $JK\Gamma$ plane is indicated in
Fig.~\ref{fig:angles}(b). Though the difference in $\Gamma'$ is still quite
large, this special point efficiently enforces the vortex-like correlations of
type \textit{a} so that the vortex phase not only grows but also becomes
dominated by \mbox{vortex-\textit{a}} pattern (c.f.
Appendix~\ref{sec:appendixPD}).
The expansion of the central zigzag phase is linked to approaching the hidden
SU(2) point that is an image of the AF Heisenberg point in
$\mathcal{T}_1\mathcal{T}_4$ transformation. This point, having
$\Gamma'\approx-0.4$ and the projection onto $JK\Gamma$ plane as indicated in
Fig.~\ref{fig:angles}(b), enforces the zigzag order with the moment direction
consistent with experiments and may be actually regarded as the source of the
central zigzag phase. On the other hand, the top zigzag region related to the
SU(2) point in the $\Gamma'=0$ plane is suppressed with $\Gamma'<0$, to the
extent that it is not even discernible already for $\GP=-0.2$.

The FM and AF phases develop more complex internal structure when $\Gamma'<0$
is added. This is due to the competition of the energy contributed by $\Gamma$
and $\Gamma'$ that are decisive for the moment direction at a classical level.
The anisotropic part of these contributions is proportional to $\pm
(\Gamma+2\Gamma')(n_x+n_y+n_z)^2$ for FM and AF, respectively. In the FM
phase, $\Gamma'<0$ creates a new subphase where the moments pushed originally
to the honeycomb plane due to $\Gamma>0$ [Fig.~\ref{fig:prob}(b)] take the
direction perpendicular to the honeycomb plane. This subphase extends near the
outer rim of the FM phase where $\Gamma$ is sufficiently weak. An opposite
effect is observed in the AF phase. Here, in addition, the absence of the
moment confinement by anisotropic classical energy in the case of
$\Gamma+2\Gamma'=0$ leads to an enhancement of quantum fluctuations and the
probability plotted in e.g. Fig.~\ref{fig:angles}(b) therefore drops at the
corresponding circle.

Based on the data presented so far, the probabilities of the best-fitting
classical configurations represent a good measure of quantum fluctuations in
the ground state. 
To have an independent quantification and to cross-check our results, we
compare them to a complementary approach, namely CMFT described in
Sec.~\ref{sec:methCMFT}. Its advantage is the ability to estimate the ordered
moment length that we plot in Fig.~\ref{fig:angles}(a). The phase boundaries
of the collinear phases are in a good agreement with the method based on ED
and the moment length reveals the less fluctuating FM and stripy phases, and
the gradual decrease of quantum fluctuations when going deeper into the AF
phase. The data on the moment angle to the honeycomb plane shows a somewhat
larger spread but the trend is identical.

The evolution of the phases with increasing third nearest-neighbor coupling
$J_3$ is illustrated in Fig.~\ref{fig:angles}(c),(d). As expected already at
the level of the classical energy, the antiferromagnetic $J_3>0$ coupling
further favors zigzag and AF phases. The stripy and vortex phases of hidden FM
nature as well as the FM phase get quickly suppressed and the two zigzag
regions merge filling the entire left half of the phase diagram. In both
zigzag and AF phases, the third nearest-neighbor bonds have contra-aligned
spins favorable for AF $J_3$ interaction. The energy gain brought by $J_3$
therefore does not visibly shift the zigzag/AF boundary. The two zigzag
regions have incompatible moment directions. When merging them, the system
makes a compromise by pushing the moment direction to the honeycomb plane so
that it can easily flip between $z$ and $(x+y)/\sqrt2$ directions projected
onto honeycomb plane [c.f. Fig.~\ref{fig:prob}(b)]. Near the boundary between
the zigzag subphases where the moment lies in the honeycomb plane, the quantum
fluctuations are significantly suppressed.

We reach the conclusion that both $\Gamma'<0$ and $J_3>0$ -- expected to be
present in real materials -- strongly stabilize the central zigzag phase that
is consistent with experimental observations in Na$_2$IrO$_3$ in both the
magnetic ordering and direction of magnetic moment.
As for the precise moment direction (figures in the right column of
Fig.~\ref{fig:angles}), the evolution seems to be dictated by $K$, $\Gamma$,
while $\GP$, $J_3$ influence mostly the extent of the phase. 
We note that one has to distinguish the real pseudospin direction and the
moment direction as probed by various techniques such as neutron or resonant
x-ray scattering \cite{Cha16}.
Based solely on the moment direction with the experimental data \cite{Chu15}
translating to the pseudospin angle of about $38^\circ-40^\circ$ \cite{Cha16},
it seems that the FM $K<0$ should be the largest interaction, followed by
possibly still large $\Gamma>0$.
Being in accord with the conclusions of Ref.~\cite{Cha16}, this also
falls in line with ab-initio estimates of dominant ferromagnetic $K$ and
comparable $J>0,\Gamma>0,\GP<0$ and $J_3>0$ \cite{Win17}.

Finally, small hatched/white areas in the zigzag phase shown in
Fig.~\ref{fig:angles} again indicate the instability of the collinear zigzag
pattern that may be interpreted as a protrusion of the possible incommensurate
phase. They appear at small $\Gamma'$ and $J_3$ which together with the
link between $\Gamma'$ and trigonal distortion suggests an explanation for the
spiral order in less distorted {\LIO} compared to {\NIO} with zigzag order.
This point was analyzed at a basic classical level in Ref.~\cite{Cha15}.



\vspace{-2mm}\section{Fluctuation-free manifolds}\label{sec:T6}\vspace{-2mm}

\begin{figure}[t]
\includegraphics[scale=1.07]{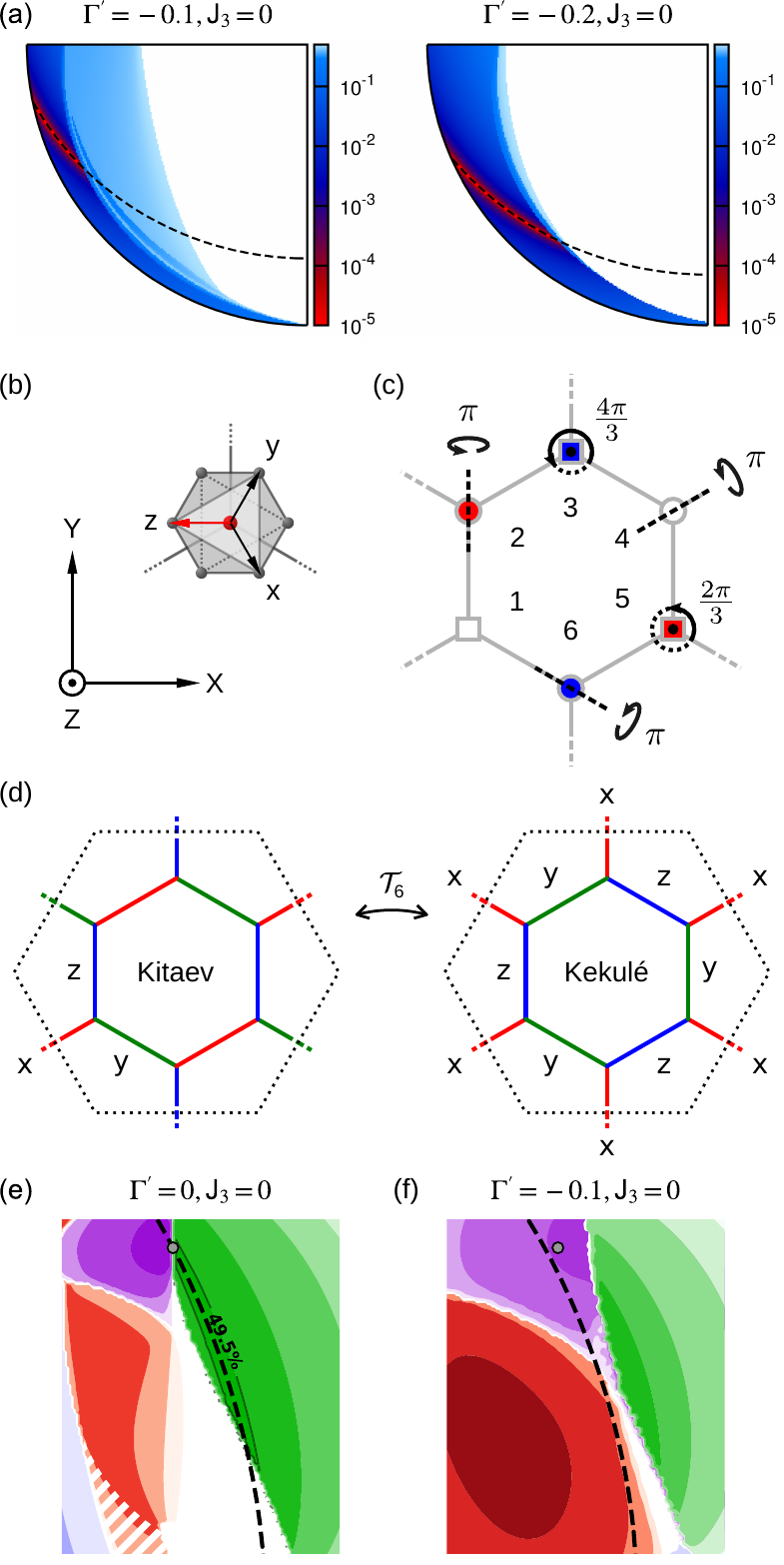}
\caption{
(a) Lower left quadrant of the phase diagram for $\Gamma'=-0.1$ and
$\Gamma'=-0.2$ showing the FM phase. The color indicates the difference
$P_\mathrm{max}-P$ between the probability $P$ of the classical state in the
ED ground state and its maximum value of $P_\mathrm{max}=\frac12$. The
maximum probability, corresponding to a fluctuation-free ground state, is
reached in the FM subphase with the moments perpendicular to the honeycomb
plane (darker color) at the line given by $K+2\Gamma-2\Gamma'=0$ (dashed).
(b) Coordinate frames used to express spin interactions. 
(c)~Schematic representation of the $\mathcal{T}_6$ transformation on the 
honeycomb lattice. At each of the six sublattices, a different rotation of
spin components is applied.
(d) Correspondence between the bonds and interaction Hamiltonians
$\mathcal{H}^{(x)}$, $\mathcal{H}^{(y)}$, $\mathcal{H}^{(z)}$ for the
EKH model and extended Kekul\'{e}-Kitaev-Heisenberg model 
obtained when performing the $\mathcal{T}_6$ transformation.
(e)~The fluctuation-free line in the AF phase of the $JK\Gamma$ model.
The line is determined by $3J+K-\Gamma-2\Gamma^\prime=0$ and
crosses the hidden SU(2) symmetric vortex point (gray~\textbullet). 
(f)~Shifted line for a case of non-zero~$\Gamma^\prime$: 
the line no longer enters the AF phase.
}
\label{fig:T6}
\end{figure}

As noticed in Sec.~\ref{sec:collin} when inspecting the phase diagram of the
$JK\Gamma$ model [Fig.~\ref{fig:prob}(a)], the AF phase contains an unusual
line of fluctuation-free ground states located near the center of the phase
diagram. The distance to this line seems to determine the magnitude of
quantum fluctuations throughout the entire AF phase -- the probability of the
N\'eel state in the ground state increases more or less monotonously starting
from the outer rim and approaching the line radially inwards. In fact, similar
lines are present also for nonzero $\Gamma'$ slices in a certain $\Gamma'$
range and form thus an entire surface in the parameter space. This is quite
unexpected since at those parameter points, all the interactions are active
and there is no apparent cancellation leading to the absence of quantum
fluctuations. What is more, a manifold of fluctuation-free ground states is
found also in part of the FM phase away from the trivially fluctuation-free FM
Heisenberg point. This is demonstrated in Fig.~\ref{fig:T6}(a) for two values
of $\Gamma'<0$. Below we address both cases, starting with the simpler FM one.

\vspace{-2mm}\subsection{FM phase}\label{sec:fffreeFM}\vspace{-2mm}

A common feature of the fluctuation-free ground states is the moment direction
being perpendicular to the honeycomb plane, suggesting to rewrite the
Hamiltonian into the $XYZ$ reference frame [Fig.~\ref{fig:T6}(b)] where this
perpendicular direction is singled out. The Hamiltonian contributions for all
the bond directions can be cast to a common form \cite{Cha15}:
\begin{align}\label{eq:Hglob}
  & \mathcal{H}^{(\gamma)}_{ij} =
  J_{XY} (S_i^X S_j^X + S_i^Y S_j^Y) + J_Z S_i^Z S_j^Z \notag \\
  & + A\,[ (S_i^X S_j^X\!-\!S_i^Y S_j^Y) \cos\phi_\gamma -
           (S_i^X S_j^Y\!+\!S_i^Y S_j^X) \sin\phi_\gamma ] \notag \\
  & - B\,[ (S_i^X S_j^Z\!+\!S_i^Z S_j^X) \cos\phi_\gamma +
           (S_i^Y S_j^Z\!+\!S_i^Z S_j^Y) \sin\phi_\gamma ] .
\end{align}
The bond-dependence of the interactions is expressed via the trigonometric
factors containing the angles of the bonds measured from the $Y$ axis, i.e.
$\phi_\gamma=0,\frac{2\pi}3,\frac{4\pi}3$ for the $c$, $a$, and $b$ bonds,
respectively. Eq.~\eqref{eq:Hglob} is obtained by inserting into
Eq.~\eqref{eq:Hcubicc} the transformation relations
\begin{equation}
\begin{pmatrix}
S^x \rule{0mm}{0mm} \\
S^y \rule{0mm}{5mm} \\
S^z \rule{0mm}{5mm}
\end{pmatrix}
=
\begin{pmatrix}
\frac1{\sqrt6}   & -\frac1{\sqrt2} & \frac1{\sqrt3} \rule{0mm}{0mm} \\
\frac1{\sqrt6}   & \frac1{\sqrt2}  & \frac1{\sqrt3} \rule{0mm}{5mm} \\
-\sqrt{\tfrac23} & 0               & \frac1{\sqrt3} \rule{0mm}{5mm}
\end{pmatrix}
\begin{pmatrix}
\phantom{-} S^X\!\cos\phi_\gamma + S^Y\!\sin\phi_\gamma \rule{0mm}{0mm} \\
         -  S^X\!\sin\phi_\gamma + S^Y\!\cos\phi_\gamma \rule{0mm}{5mm} \\
            S^Z                                         \rule{0mm}{5mm}
\end{pmatrix}
\end{equation}
which represent a conversion from the cubic $xyz$ to $XYZ$ reference frame for a $c$ bond as
well as the necessary cyclic permutation among $xyz$ (rotation around $Z$
axis), and using the fact that $\cos2\phi_\gamma=\cos\phi_\gamma$,
$\sin2\phi_\gamma=-\sin\phi_\gamma$ for the allowed values of $\phi_\gamma$.

The interaction parameters in \eqref{eq:Hglob} expressed in
terms of the original $J$, $K$, $\Gamma$, and $\Gamma'$ read as
\begin{align}
  J_{XY} &= J+\tfrac13(K-\Gamma-2\Gamma') , \\
  J_Z    &= J+\tfrac13(K+2\Gamma+4\Gamma') , \\
  A      &= \tfrac13 (K+2\Gamma-2\Gamma') , \\
  B      &= \tfrac{\sqrt{2}}3 (K-\Gamma+\Gamma') . \label{eq:Hglobpars} 
\end{align}
Let us now consider a FM state polarized in the $Z$ direction and inspect the
terms that could lead to quantum fluctuations. As in usual Heisenberg
magnets, the $J_{XY}$ interaction containing $S_i^+ S_j^-$ and $S_i^- S_j^+$
does not act on the polarized state. The above state is an eigenstate of the
$S_i^Z$ operators, the action of $B$-terms in the Hamiltonian therefore sums
up to 
\begin{equation}
  -B \sum_\mathrm{sites}\left[
    S_i^X \sum_{\gamma=a,b,c}\!\! \cos\phi_\gamma
   +S_i^Y \sum_{\gamma=a,b,c}\!\! \sin\phi_\gamma \right]
\end{equation}
which drops out since both $\sum_\gamma\cos\phi_\gamma$ and
$\sum_\gamma\sin\phi_\gamma$ are zero. Only the remaining $A$-terms containing
$S_i^+ S_j^+$ and $S_i^- S_j^-$ are active. Setting $A=0$, all the $S_i^-$ or
$S_i^-S_j^-$ terms that could lead to quantum fluctuations are cut off by zero
prefactors and we are left with an exact eigenstate. The two conditions for a
fluctuation-free FM ground state, i.e. moments being perpendicular to the
honeycomb plane and $A=0$ translating to 
\begin{equation}\label{eq:ffcondFM}
  K+2\Gamma-2\Gamma'=0 ,
\end{equation}
are checked in Fig.~\ref{fig:T6}(a). Approaching the line given by
Eq.~\eqref{eq:ffcondFM} within the $(111)$ polarized FM phase, the probability
indeed reaches the maximum value of $\frac12$, reflecting the two degenerate
configurations (moments along $Z$ or $-Z$) being superposed in the cluster
ground state. Exactly at this line, we find a doubly degenerate ground state.
Finally, let us note that the ground state remains fluctuation-free even in
the presence of $J_3$ provided that the $(111)$ polarized FM pattern is
preserved.

\vspace{-2mm}\subsection{AF phase}\vspace{-2mm}

A more complex situation is encountered in the case of $(111)$ polarized AF
phase. Here the fluctuation-free manifold is attached to the vortex point of
hidden SU(2) symmetry hosting an infinite number of fluctuation-free states.
The $(111)$ polarized AF state is one of them, the others being e.g. the
\mbox{vortex-\textit{a}} and \mbox{vortex-\textit{b}} configurations shown in
Fig.~\ref{fig:prob}(b). The connection to the SU(2) vortex point suggests a
special role of the $\mathcal{T}_6$ transformation which we discuss in more
detail here.

The $\mathcal{T}_6$ transformation is a 6-sublattice mapping 
that rotates the spins according to the recipe:
\begin{align}\label{eq:T6array}
  \text{sublattice 1:} \quad 
   (S^{x\prime},S^{y\prime},S^{z\prime}) &= ( S^x, S^y, S^z) , \notag \\
  \text{sublattice 2:} \quad 
   (S^{x\prime},S^{y\prime},S^{z\prime}) &= (-S^y,-S^x,-S^z) , \notag \\ 
  \text{sublattice 3:} \quad 
   (S^{x\prime},S^{y\prime},S^{z\prime}) &= ( S^y, S^z, S^x) , \notag \\ 
  \text{sublattice 4:} \quad 
   (S^{x\prime},S^{y\prime},S^{z\prime}) &= (-S^x,-S^z,-S^y) , \notag \\ 
  \text{sublattice 5:} \quad 
   (S^{x\prime},S^{y\prime},S^{z\prime}) &= ( S^z, S^x, S^y) , \notag \\ 
  \text{sublattice 6:} \quad 
   (S^{x\prime},S^{y\prime},S^{z\prime}) &= (-S^z,-S^y,-S^x) .
\end{align}
For a better understanding, the transformation is depicted in
Fig.~\ref{fig:T6}(c). On sites $1$, $3$, $5$ marked by a square symbol, the
spins are rotated around the $(111)$ axis, on sites $2$, $4$, $6$ marked by a
circle, the mapping consists of $\pi$-rotations around axes lying in the
honeycomb plane. This in effect changes the $(111)$ polarized AF pattern into
$(111)$ polarized FM one, making a first step towards the understanding of the
AF fluctuation-free line.

The second step involves the transformation of the Hamiltonian. Performing the
$\mathcal{T}_6$ spin rotations, we find that the resulting model is similar to
EKH in the sense that three types of bond interactions of the form of
Eq.~\eqref{eq:Hcubicc} appear,
$\mathcal{H}^{(x)}\equiv\mathcal{H}^{(a)}$,
$\mathcal{H}^{(y)}\equiv\mathcal{H}^{(b)}$, and
$\mathcal{H}^{(z)}\equiv\mathcal{H}^{(c)}$,
with the parameters modified according to
\begin{equation}\label{eq:Kekule}
  (J,K,\Gamma,\Gamma')_\text{Kekul\'{e}} = (-\Gamma,-J-K+\Gamma,-J,-\Gamma') .
\end{equation}
However, the assignment of $\mathcal{H}^{(x,y,z)}$ to the bonds is not simply
by the bond direction anymore. Instead, as shown in Fig.~\ref{fig:T6}(d), a
network of benzene-like rings governed by alternating $\mathcal{H}^{(y)}$ and
$\mathcal{H}^{(z)}$ is formed. They are interconnected by bonds possessing the
$\mathcal{H}^{(x)}$ type of interactions. This way, the $\mathcal{T}_6$
transformation maps the EKH model to an extended variant of Kekul\'{e}--Kitaev
model \cite{Qui15}.

We are now in position to combine the result of $\mathcal{T}_6$ transformation
with the argumentation of Sec.~\ref{sec:fffreeFM}. Since the transformation
led to $(111)$ polarized FM pattern and in the new model each site is a member
of three bonds governed by $\mathcal{H}^{(x)}$, $\mathcal{H}^{(y)}$, and
$\mathcal{H}^{(z)}$, the cancellation of the terms leading to quantum
fluctuations proceeds exactly the same way. Substituting the parameters in
Eq.~\eqref{eq:ffcondFM} according to \eqref{eq:Kekule}, we thus arrive at the
condition for the fluctuation-free AF state:
\begin{equation}\label{eq:ffcondAF}
  3J+K-\Gamma-2\Gamma^\prime=0 .
\end{equation}
As demonstrated in Fig.~\ref{fig:T6}(e), this line coincides with the region
where the probability of N\'eel state peaks at $\frac 12$. For a negative
$\Gamma'$, the line quickly gets out of the AF phase. However, going in the
positive $\Gamma'$ direction, the fluctuation-free line gets even deeper into
the AF phase (c.f. Appendix~\ref{sec:appendixPD}). At the special point
$J=\Gamma=\Gamma'>0$, $K=0$ on the fluctuation-free manifold, the model even
reduces to AF Ising model with the $(111)$ Ising axis, as can be seen from
Eqs.~\eqref{eq:Hglob}-\eqref{eq:Hglobpars}.

Unlike in the previous FM case, the addition of $J_3$ spoils the
fluctuation-free nature of the ground state since the $J_3$ interaction
generates terms of $A$-type under the $\mathcal{T}_6$ transformation.


\vspace{-2mm}\section{Ising-Kitaev-compass model}\label{sec:ising}\vspace{-2mm}

In this section, we address yet another feature of the model that enables
further insights into its phase behavior. Namely, we find points in the
parameter space where the four interactions $JK\Gamma\Gamma'$ can be combined
into a single one, characterized by a single interacting spin component
(interaction axis) that depends on the bond direction. This way, the model in
Eq.~\eqref{eq:Hglob} may realize combinations of Ising, Kitaev, or compass
model on the honeycomb lattice.

\vspace{-2mm}\subsection{Compass point in the phase diagram}\vspace{-2mm}
\label{sec:kitaevisingpoint}

\begin{figure}[t!]
\includegraphics[scale=1.0]{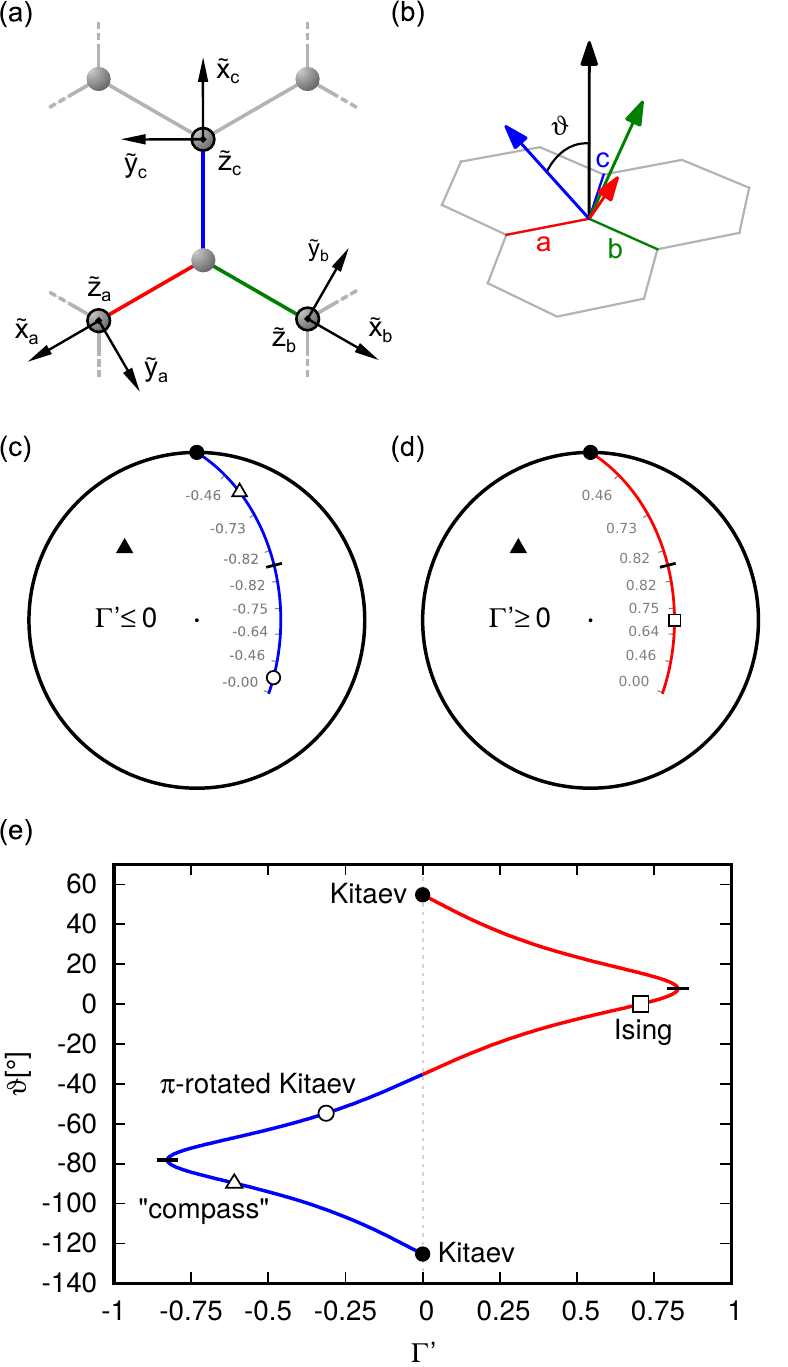}
\caption{
(a)~Rotating coordinate system
$\tilde{x}_\gamma,\tilde{y}_\gamma,\tilde{z}_\gamma$ on a honeycomb lattice --
color distinguishes three bond types $a$, $b$, $c$.
(b)~The direction of the interaction axis for each bond. All of them are at an
angle $\vartheta$ with the Ising $(111)$ direction shown in black.
(c),(d)~Ising-Kitaev-compass parameter line in the phase diagram and the
corresponding values of the $\GP$ parameter.  The Ising point ($\Box$) emerges
for positive $\GP$, while the $\pi$-rotated Kitaev point ($\circ$) and the
perpendicular ``compass'' point ($\vartriangle$) are found for negative $\GP$
values. The true compass point ($\blacktriangle$) appears for $\GP=0$.
(e)~The angle $\vartheta$ of the interaction axis to the $(111)$ direction
depending on the position on the parameter line. $\varphi=\pi/2$ is assumed.
The Kitaev point (\textbullet) is connected to the $\pi$-rotated Kitaev point
($\circ$) by the $\mathcal{T}_1$ dual transformation \cite{Cha15} -- 
a $\pi$-rotation around the $(111)$ axis.
}
\label{fig:kitaevising}
\end{figure}

Inspecting the $JK\Gamma$ phase diagram, we find a degenerate point, in which
several phases seem to meet: vortex, ferromagnet, and both zigzag phases.
Writing the interaction as $\mathcal{H}=\sum_{ij}\vc{S}^T_iH_{ij}\vc{S}_j$, we
find that the Hamiltonian matrices in this parameter point $K=\Gamma=-J >0$,
$\GP=0$ have a~symmetrical block shape:
\begin{align}
  H_a&=
  \begin{pmatrix}
    J+K & \GP & \GP \\
    \GP & J & \Gamma \\ 
    \GP & \Gamma & J
  \end{pmatrix}=
  \begin{pmatrix}
    0 & 0 & 0 \\
    0 & -K & K \\ 
    0 & K & -K
  \end{pmatrix},\\  
  H_b&=
  \begin{pmatrix}
    J & \GP & \Gamma \\
    \GP & J+K & \GP \\ 
    \Gamma & \GP & J
  \end{pmatrix}=
  \begin{pmatrix}
    -K & 0 & K \\
    0 & 0 & 0 \\ 
    K & 0 & -K
  \end{pmatrix},\\ 
  H_c&=
  \begin{pmatrix}
    J & \Gamma & \GP \\
    \Gamma & J & \GP \\ 
    \GP & \GP & J+K
  \end{pmatrix}=
  \begin{pmatrix}
    -K & K & 0 \\
    K & -K & 0 \\ 
    0 & 0 & 0
  \end{pmatrix}.
\end{align}
The matrices can be diagonalized by a change of basis to the rotating
coordinate frame $\tilde{x}_\gamma,\tilde{y}_\gamma,\tilde{z}_\gamma,
\gamma\in\{a,b,c\}$, where $\tilde{x}_\gamma$ axis points in the bond
direction, $\tilde{y}_\gamma$ is perpendicular to the bond direction and lies
in the honeycomb plane, and $\tilde{z}_\gamma$ points out of the honeycomb
plane -- see Fig. \ref{fig:kitaevising}(a) for a sketch of this coordinate
system. After the change of basis, all three interaction matrices have the
same form for all bond directions $a$, $b$, $c$:
\begin{equation}
  \widetilde{H}=
  \begin{pmatrix}
  -2K & 0 & 0 \\
    0 & 0 & 0 \\
    0 & 0 & 0
  \end{pmatrix},
\end{equation}
which represents FM interaction in the bond direction, concisely written as:
\begin{equation}
  \mathcal{H}=\sum_{\langle ij\rangle\in\mathrm{NN}}
  -2K(\vc{S}_i\cdot\vc{r}_{ij})(\vc{S}_j\cdot\vc{r}_{ij}),
\end{equation}
where the unit vector $\vc{r}_{ij}$ points from site $i$ to site $j$. This
form of interaction is known in the literature as the 120$^{\circ}$ honeycomb
compass model \cite{Zha08,Wu08,Nas08,Nus15}. Similar to the Kitaev model, it
features frustration due to competing interactions for the three bond
directions. However, the exact ground state is not known in this case and its
nature is in fact not clear, as several past works came to inconsistent
conclusions.  One study found a N\'{e}el state \cite{Zha08}, others suggested
a stabilization of a dimer pattern \cite{Wu08}, a superposition of dimer
coverings \cite{Nas08}, or a quantum spin liquid state \cite{Che16}.  With the
link to the compass model, the apparent special role of the $-J=K=\Gamma$
point marked by a competition of four long-range ordered phases in its
vicinity is confirmed. This competition was noticed also in the
tensor-network analysis of the $JK\Gamma$ model \cite{Lou15}, which claimed a
small region surrounding this point to harbor a valence bond solid phase. 

\vspace{-2mm}\subsection{Ising-Kitaev-compass line in the phase diagram}\vspace{-2mm}

Motivated by the previous example, we now demonstrate that the EKH model
provides also a more general case of a single interaction axis lying anywhere
between the honeycomb plane and the perpendicular $(111)$ direction. Dictated
by the $C_3$ symmetry of the EKH model, the interaction axis has to rotate
together with the bond direction as shown in Fig.~\ref{fig:kitaevising}(b).
Representing the bond-dependent interaction axis by a unit vector 
$\vc n_\gamma$, the Hamiltonian of the single-axis model hidden in 
the EKH model has to be of the form 
\begin{equation}\label{eq:kitaevising_gen}
  \mathcal{H}_\mathrm{IKc}
  =\sum_{\langle ij\rangle\in\mathrm{NN}}\widetilde{K}
  (\vc{n}_\gamma\cdot\vc{S}_i)(\vc{n}_\gamma\cdot\vc{S}_j) .
\end{equation}
We specify the interaction-axis direction by the deviation $\vartheta$ from
the $(111)$ direction and the azimuthal angle $\varphi$ measured from the
bond. In the rotating reference frame
$\tilde{x}_\gamma\tilde{y}_\gamma\tilde{z}_\gamma$ we have
\begin{equation}\label{eq:kitaevisingaxis}
  \vc{n}_\gamma=\bigl(\,
  \sin\vartheta\cos\varphi,
  \sin\vartheta\sin\varphi,
  \cos\vartheta \,\bigl) 
\end{equation}
while in the $XYZ$ reference frame of Fig.~\ref{fig:T6}(b)
\begin{equation}
  \vc{n}_\gamma=\bigl(\,
  -\sin\vartheta\sin(\varphi\!+\!\phi_\gamma),
  \sin\vartheta\cos(\varphi\!+\!\phi_\gamma),
  \cos\vartheta \,\bigl) ,
\end{equation}
with $\phi_\gamma$ being the bond angles defined in Sec.~\ref{sec:T6}. In the
direction of increasing $\vartheta$, the Hamiltonian
\eqref{eq:kitaevising_gen} encompasses Ising model ($\vartheta=0$), Kitaev
model ($\vartheta=\arccos(1/\sqrt3)$, $\varphi=\pi/2$), and compass model
($\vartheta=\pi/2$, $\varphi=0$).
Note, that the physical properties do not depend on $\varphi$ so that it is
sufficient to focus on $\vartheta$ as the relevant parameter. For example,
true compass model has $\varphi=0$ (the interaction axis coincides with the
bond direction) but $\varphi=\pi/2$ (the in-plane interaction axis is
perpendicular to the bonds) leads to the same ground state, apart from a
trivial rotation. We thus call the latter one ``compass'' to suggest a small
only distinction.

We now establish the link to EKH model by converting its interactions into
main axes. This is most conveniently performed in the rotating reference frame
where the Hamiltonian is represented by a matrix common to all bond
directions. If two of its eigenvalues are zero, we are left with the single
interaction axis corresponding to the model of Eq.~\eqref{eq:kitaevising_gen}.
In Sec.~\ref{sec:kitaevisingpoint} we have already encountered a situation in
which the Hamiltonian was readily diagonalized merely by casting it into the
rotating frame and the only nonzero eigenvalue for the in-bond direction
generated the compass interaction. The general inspection is left for
Appendix~\ref{sec:appendixIKc}, here we only summarize the results presented
in Fig.~\ref{fig:kitaevising}(c)-(e).
It turns out that the compass case with $\vc n_\gamma = \tilde{x}_\gamma$
($\varphi=0$) is singular for $\Gamma\geq 0$ and in the other cases the
interaction axis is found in the perpendicular
$\tilde{y}_\gamma\tilde{z}_\gamma$ plane ($\varphi=\pi/2$) --
Fig.~\ref{fig:kitaevising}(b) contains a sketch of such a bond-dependent
interaction axis.

Assuming $\Gamma>0$, the diagonalized interaction has only one non-zero
component on the line determined by
\begin{equation}\label{eq:kitaevisingline}
  J=\Gamma,\;J(J+K)=\Gamma^{\prime 2},\;J>0,\;J+K>0
\end{equation}
which is indicated in Fig. \ref{fig:kitaevising}(c),(d). In our
parametrization, it covers the range 
$|\Gamma'|\leq \tfrac12 \sqrt{1+\sqrt3}\approx 0.83$. 
Fig.~\ref{fig:kitaevising}(e) shows how the deviation $\vartheta$ of the
interaction axis from $(111)$ direction evolves on the parameter line
\eqref{eq:kitaevisingline}; the colors differentiate the two branches for
$\GP>0$ and $\GP<0$, respectively, and correspond to colors in
Fig.~\ref{fig:kitaevising}(c),(d). The interaction constant $\tilde{K}=3J+K$
is always positive, hence the interaction is antiferromagnetic. 
Several distinct points are labeled in the figure: The limit $\vartheta=0$
achieved for $J=\Gamma=\GP=\tfrac13\tilde{K}$, $K=0$ corresponds to an
antiferromagnetic Ising point that lies at the same time at the
fluctuation-free manifold discussed in Sec.~\ref{sec:T6}. The ``compass''
limit $\vartheta=\pm\pi/2$ is reached for $J=\Gamma=\tfrac16\tilde{K}$,
$K=\tfrac12\tilde{K}$, $\Gamma'=-\tfrac13\tilde{K}$.  By varying the model
parameters, one can arbitrarily interpolate between these two limits. A
special role is played by the Kitaev case $\vartheta\approx \pm 54.7^\circ$
that is characterized by a spin-liquid ground state. It can be found either at
AF Kitaev point $K=\tilde{K}$, $J=\Gamma=\GP=0$ with
$\vartheta=\arccos(1/\sqrt3)$ or, less trivially, in a form rotated by $\pi$
about the $(111)$ axis: $J=\Gamma=\tfrac49\tilde{K}$, $K=-\tfrac13\tilde{K}$,
$\Gamma'=-\tfrac29\tilde{K}$ with $\vartheta=-\arccos(1/\sqrt3)$. In general,
the parameters for $\vartheta$ and $-\vartheta$ are related by the
$\mathcal{T}_1$ transformation of Ref.~\cite{Cha15} that corresponds to
a $\pi$-rotation about the $(111)$ axis (for details see
Appendix~\ref{sec:appendixIKc}).


\vspace{-2mm}\subsection{Phases of Ising-Kitaev-compass model}\vspace{-2mm}

Having identified the line in the phase diagram where the EKH model
effectively interpolates between Ising, Kitaev, and ``compass'' model captured
by Eq.~\eqref{eq:kitaevising_gen} (all AF for $\Gamma>0$), we now briefly
address the phase diagram on this line for varying $\vartheta$. Let us note,
that the corresponding type of model (dubbed ``tripod'') has been studied
before using tensor networks \cite{Zou16}, although with the axis in the
$\tilde{x}_\gamma\tilde{z}_\gamma$ plane and not in connection with the EKH
model. As before, we apply the method of Sec.~\ref{sec:methcoh} combined with
an analysis of the spin correlations.

\begin{figure}[t]
\includegraphics[scale=1.02]{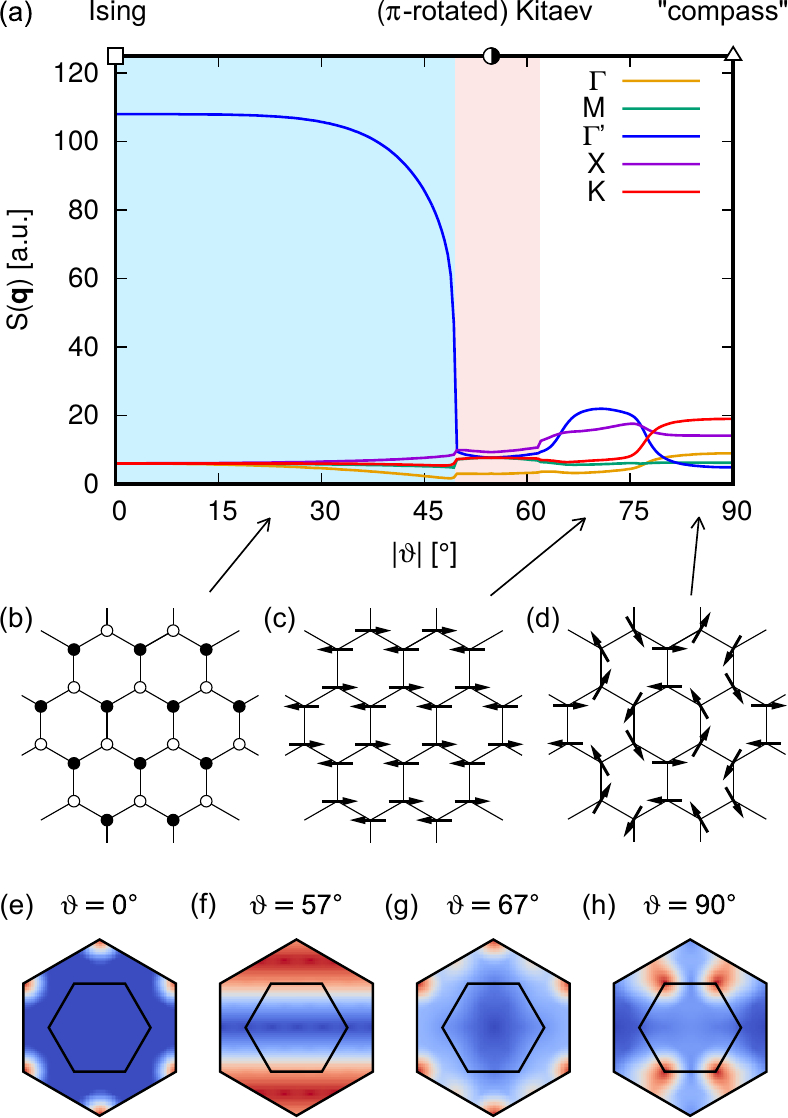}
\caption{
(a)~Trace of the spin structure factor 
$S(\vc{q})= \sum_\alpha\langle
S^\alpha_{-\vc{q}}S^\alpha_{\vc{q}}\rangle$
obtained by ED using hexagonal $24$-site cluster.  The calculation reveals an
Ising antiferromagnet and a Kitaev liquid phase; for high values of
$|\vartheta|$ (interaction axis near the honeycomb plane), the correlations
show a mixture of contributions mainly at $\GP$, X and K points. 
(b)-(d)~Most probable spin configurations determined by the method of
spin-$\tfrac{1}{2}$ coherent states (using $\varphi=\pi/2$). The N\'{e}el
state is out of plane in (b) and only slightly tilted from the plane in (c).
The spin arrangement (d) is the same as the \mbox{vortex-$a$} configuration 
shown in Fig.~\ref{fig:prob}(b).
(e)-(h)~Spin-spin correlations $\langle S^z_{-\vc{q}}S^z_{\vc{q}}\rangle$ for
a few selected values of $\vartheta$ and $\varphi=\pi/2$. In panel (f), the
typical cosine wave pattern characterizing a spin liquid state appears.
}
\label{fig:kitaevisingsqsq}
\end{figure}

The resulting phase diagram is shown in Fig.~\ref{fig:kitaevisingsqsq}.  We
formally plot it as function of $|\vartheta|$.  For negative values of
$\vartheta$ it covers the range Ising -- $\pi$-rotated Kitaev -- ``compass"
that is continuously visible in Fig.~\ref{fig:kitaevising}(e). The phase
diagram for positive $\vartheta$ is identical, even though corresponding to
different line in the parameter space of the EKH model.

More than half of the phase diagram is occupied by the AF Ising phase with the
moments pointing in the $(111)$ direction. It starts as fluctuation free at
the Ising point $\vartheta=0^\circ$ and gradually acquires more quantum
fluctuating nature until $|\vartheta|\approx 49.5^\circ$ where a transition to a
spin-liquid associated with the ($\pi$-rotated) Kitaev point at
$|\vartheta|\approx 54.7^\circ$ happens. The increasing content of quantum
fluctuations is manifested by decreasing spin correlations
[Fig.~\ref{fig:kitaevisingsqsq}(a)] or the probability of N\'eel state in the
ground state which follows a similar curve. At approximately $62^\circ$ the
spin liquid state breaks down. Compared to the tensor network study
\cite{Zou16}, we find a significantly wider spin liquid window --
$(49.5^\circ,62^\circ)$ vs $(52.7^\circ,57.6^\circ)$. The reason for this
discrepancy is not clear, an earlier tensor-network study of the
Kitaev-Heisenberg model \cite{Ire14} was in a good agreement with ED. One
possibility is that a~different procedure of finding the level crossings or
extrapolating the bond dimension has been used for the two tensor networks
studies. 

For $|\vartheta|\gtrsim 62^\circ$, the ground state is characterized by peaks
in the structure factor at the $\GP$ and X point [see
Fig.~\ref{fig:kitaevisingsqsq}(g)], while around $|\vartheta|\approx 77^\circ$
the correlations at $\GP$ point drop and the K point becomes dominant
[Fig.~\ref{fig:kitaevisingsqsq}(h)]. 
In the first part of this interval, the method of spin-$\tfrac{1}{2}$ coherent
states finds a N\'eel state with the moment direction slightly tilted
($\lesssim 5^\circ$) from the honeycomb plane
[Fig.~\ref{fig:kitaevisingsqsq}(c)] and the amount of quantum fluctuations
comparable to the ground state of the AF Heisenberg model. 
After a transition at about $77^\circ$, a \mbox{vortex-\textit{a}} pattern is 
found [Fig.~\ref{fig:kitaevisingsqsq}(d)], but with a probability 
$P\approx 3\%$ significantly smaller than observed earlier inside the vortex 
phase. This suggests that the second phase might be possibly disordered, as 
claimed previously by \cite{Nas08} and \cite{Che16} for the compass model. 
The absence of visible changes in the correlations in the region of the second
phase indicates that the features of ``compass'' limit $|\vartheta|=90^\circ$ 
are kept throughout this phase. 
Similarly to the vortex phase, the most probable pattern
(\mbox{vortex-\textit{a}}) is accompanied by the complementary pattern
(\mbox{vortex-\textit{b}}) having a very close probability ($P\approx 3.16\%$
vs $3.05\%$ in the AF ``compass'' limit). The same pair but with the swapped
probabilities is found for FM compass point discussed in
Sec.~\ref{sec:kitaevisingpoint}. This is natural since the two models as well
as the two patterns are linked by a simple $90^\circ$ rotation of the spins
(interchanging the in-bond and perpendicular components), followed by a
$180^\circ$ rotation at every second site (converting AF to FM and vice
versa).

Finally, we note that the presence of two distinct phases between the spin
liquid phase and the compass limit is at odds with the conclusion of
Ref.~\cite{Zou16} that the whole interval is occupied by a dimer phase. 
To check the reliability of our phase diagram, we have performed an additional
ED for $32$-site clusters of two different shapes, confirming the existence of
the N\'{e}el phase for those clusters as well.



\vspace{-2mm}\section{Conclusions}\vspace{-2mm}

We have performed a detailed numerical investigation of the global
phase diagram of the extended Kitaev-Heisenberg model including the analysis
of the internal structure of the individual phases. To this end, we have used
mainly exact diagonalization combined with a recently developed ground-state
analysis based on spin-$\frac12$ coherent states.

In the context of real materials such as {\NIO} or {\RuC}, our results are
useful when judging the extent of the experimentally observed zigzag phase and
comparing the direction of the ordered moments, fixing thereby the relevant
window in the parameter space.

In more general terms, we have interpreted the trends in the phase diagram
based on several types of symmetry features found in the extended
Kitaev-Heisenberg model, providing a number of reference points of expected
behavior. They include points of hidden SU(2) symmetry, manifolds of
fluctuation-free ground states, and mappings to other models: Ising, compass,
and hidden Kitaev model. We have demonstrated that while being in principle
simple results of linear algebra, these symmetry features have far-reaching
consequences and well fix the overall structure of the global phase diagram.
As we believe, our symmetry-guided study can serve as a methodological
template that can be applied to other spin models with bond-dependent
non-Heisenberg interactions emerging in the field of Mott insulators with
strong spin-orbit coupling.

Among the unusual symmetry properties brought about by the bond-dependent
non-Heisenberg interactions, we have highlighted two interesting features
that, to the best of our knowledge, escaped attention so far: \\
(i)~Fluctuation-free ground states on entire manifolds of parameter points,
possessing both FM and AF ordering patterns. These are enabled due to a
partial cancellation of interactions for the particular spin structure.
However, above the ground state, these interactions are fully active and shall
lead to excitation spectra quite distinct from those of e.g. Heisenberg FM. \\
(ii)~Models with bond-dependent non-Heisenberg interactions may realize not
only the sought-after Kitaev model but also other models with a single
bond-dependent interacting spin component (``interaction axis'').  In the case
of the extended Kitaev-Heisenberg model, they range from the simple Ising
model, through Kitaev, to the $120^\circ$ compass model on honeycomb lattice,
whose ground-state nature is still debated in the literature. The above models
are continuously connected in the parameter space of the extended
Kitaev-Heisenberg model and the corresponding line in the phase diagram
contains both trivial (Ising limit) as well as highly nontrivial phases --
perturbed Kitaev spin liquid and the phase associated with the perturbed
compass model.
Such links to the extended Kitaev-Heisenberg model may motivate a
search for candidate materials realizing, e.g. a compass model.


\acknowledgments

We would like to thank Giniyat Khaliullin, Andrzej M. Ole\'{s}, Krzysztof
Wohlfeld, and Kurt Hingerl for helpful discussions. 
JC and JR acknowledge support from the Ministry of Education, Youth and Sports
of the Czech Republic within the project CEITEC 2020 (LQ1601) under the
National Sustainability Programme II, European Research Council via project
TWINFUSYON (No.~692034), Czech Science Foundation (GA\v{C}R) under project No.
GJ15-14523Y, and Masaryk University internal projects MUNI/A/1310/2016 and
MUNI/A/1291/2017. JR is Brno Ph.D Talent Scholarship Holder -- Funded by the
Brno City Municipality. 
DG was supported by Polish National Science Center (NCN) under projects
2012/04/A/ST3/00331 and 2016/23/B/ST3/00839. 
Computational resources were provided by the CESNET LM2015042 and the CERIT
Scientific Cloud LM2015085, provided under the programme ``Projects of Large
Research, Development, and Innovations Infrastructures''. The CMFT
calculations were performed at the Interdisciplinary Centre for Mathematical
and Computational Modelling (ICM) of the University of Warsaw under grants 
No.~\mbox{G66-22} and \mbox{G72-9}. 


\appendix

\vspace{-2mm}\section{Optimized spin-$\frac12$ coherent states in the presence of spin liquid phases}\vspace{-2mm}
\label{sec:appendixSL}

\begin{figure}[t]
\includegraphics[scale=1.0]{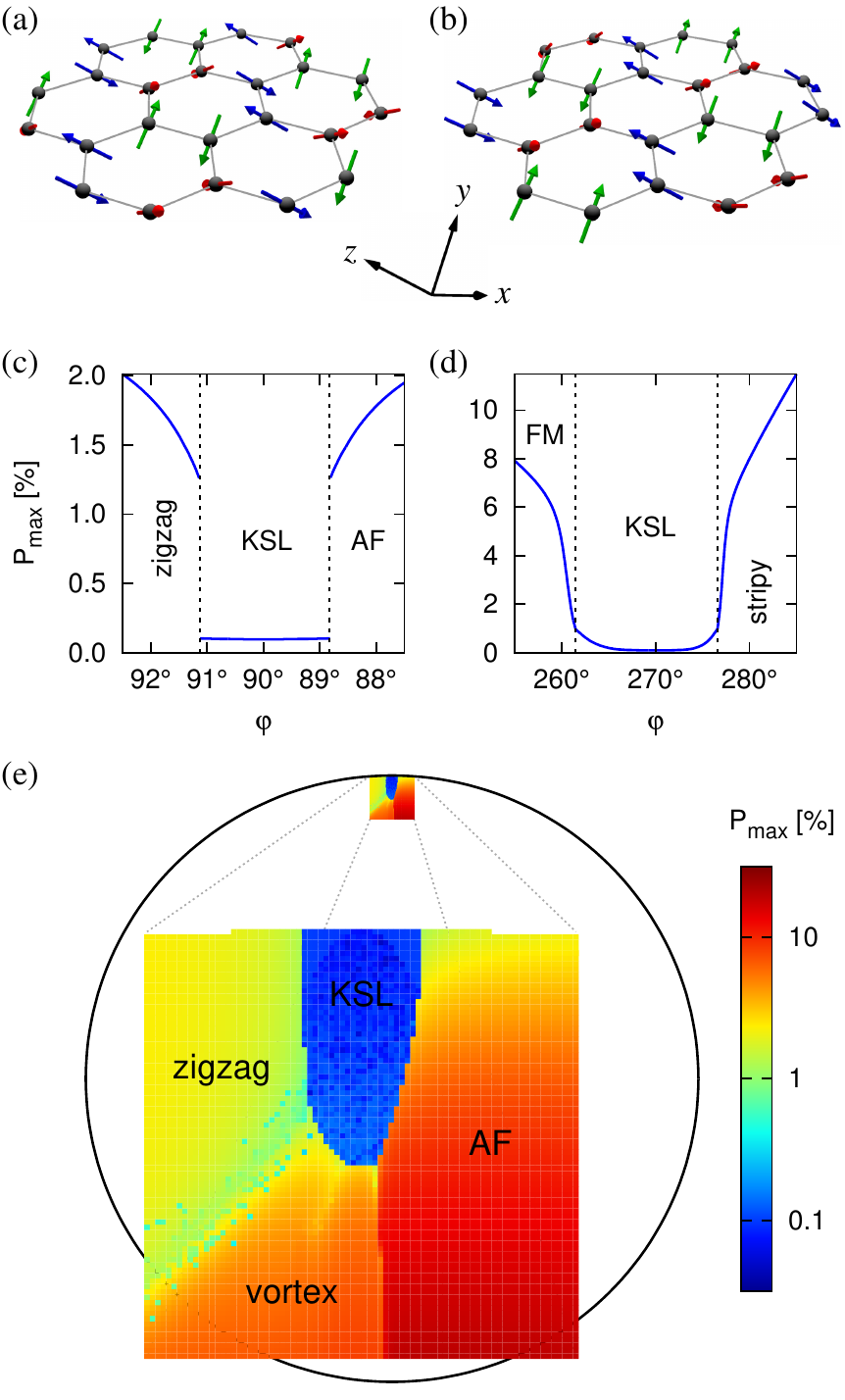}
\caption{
Most probable configurations near the (a) AF and (b) FM Kitaev point. Spins
pointing along the $x$, $y$, and $z$ cubic axes are marked by red, green, and
blue color, respectively.
(c) Probability $P_\mathrm{max}$ of the optimized configuration near the AF
Kitaev point in the Kitaev-Heisenberg model parametrized as $J=\cos\varphi$,
$K=\sin\varphi$. Phase transitions to the neighboring ordered phases are
indicated by dashed lines. (d) The same for the region around the FM Kitaev
point. 
(e) Map of $P_\mathrm{max}$ in a small area near the AF Kitaev spin liquid
phase in the extended Kitaev-Heisenberg model with $\Gamma'=J_3=0$ (belongs to
the same slice as shown in Fig.~\ref{fig:prob}). 
}
\label{fig:SL}
\end{figure}

In contrast to the ordered phases, the ground states of Kitaev spin liquid
(KSL) phases are highly entangled and cannot be well described by a
spin-$\frac12$ coherent state \eqref{eq:cohstate} that is a simple product of
spin-$\frac12$ states. This is indeed observed when applying the method of
Sec.~\ref{sec:methcoh} to the spin-liquid ground state. Nevertheless, the
spin-liquid ground state still has a significant overlap with the
configurations found in the classical $S\rightarrow \infty$ limit of the
Kitaev model. 
Shown in Fig.~\ref{fig:SL}(a),(b) are the most probable configurations found
in the exact ground state of the 24-site cluster near the AF or FM Kitaev
point, respectively. They are characterized by spins pointing along the cubic
axes $x$, $y$, $z$ and forming aligned (FM case) or contra-aligned (AF case)
pairs on nearest-neighbor bonds. Their orientation is determined by the bond
direction which is linked to the active spin component in the Kitaev
interaction.

It is instructive to inspect the evolution of probability $P_\mathrm{max}$ of
the optimized spin-$\frac12$ coherent state when going from the KSL phase
through a quantum phase transition to the neighboring ordered phases. This is
done in Fig.~\ref{fig:SL}(c) for the case of AF Kitaev model perturbed by
Heisenberg interaction. While the KSL is characterized by small
$P_\mathrm{max}\approx 0.1\%$, at the phase boundary either to zigzag or to AF
ordered phase $P_\mathrm{max}$ jumps by about one order of magnitude to values
above $1\%$ typically encountered in the main text for the ordered phases with
significant quantum fluctuations.
A different behavior is found in the FM case. Here the probability rises near
the phase boundaries to FM and stripy phases and continuously connects to
$P_\mathrm{max}$ of their groundstates. However, a discontinuity still appears
in the derivative with respect to model parameters, with $P_\mathrm{max}$
shooting up after crossing the phase boundary. This is yet another
manifestation of the different nature of the phase transitions involving AF
and FM KSL phases that shows up e.g. in the behavior of the spin-excitation
gap \cite{Got17}.

Finally, in Fig.~\ref{fig:SL}(e) we consider a larger portion of the phase
diagram near the AF KSL phase which also involves the nonzero $\Gamma>0$
parameter case. The AF KSL phase can be easily distinguished again by a drop
of $P_\mathrm{max}$. The case of FM KSL phase at the bottom of the
corresponding phase diagram slice is more complicated by the less pronounced
transition and more complex phase behavior in the surrounding region as
discussed in Sec.~\ref{sec:white}. Fig.~\ref{fig:SL}(e) also illustrates the
difficulties of the global optimization in the complete 48-dimensional space
of $\theta$ and $\phi$ parameters corresponding to the 24-site cluster. As
seen in Fig.~\ref{fig:SL}(e), the danger of getting trapped in a local maximum
increases e.g. near the zigzag/vortex phase boundary or in the KSL phase
characterized by competing configurations.


\begin{figure*}[t]
\includegraphics[scale=1.01]{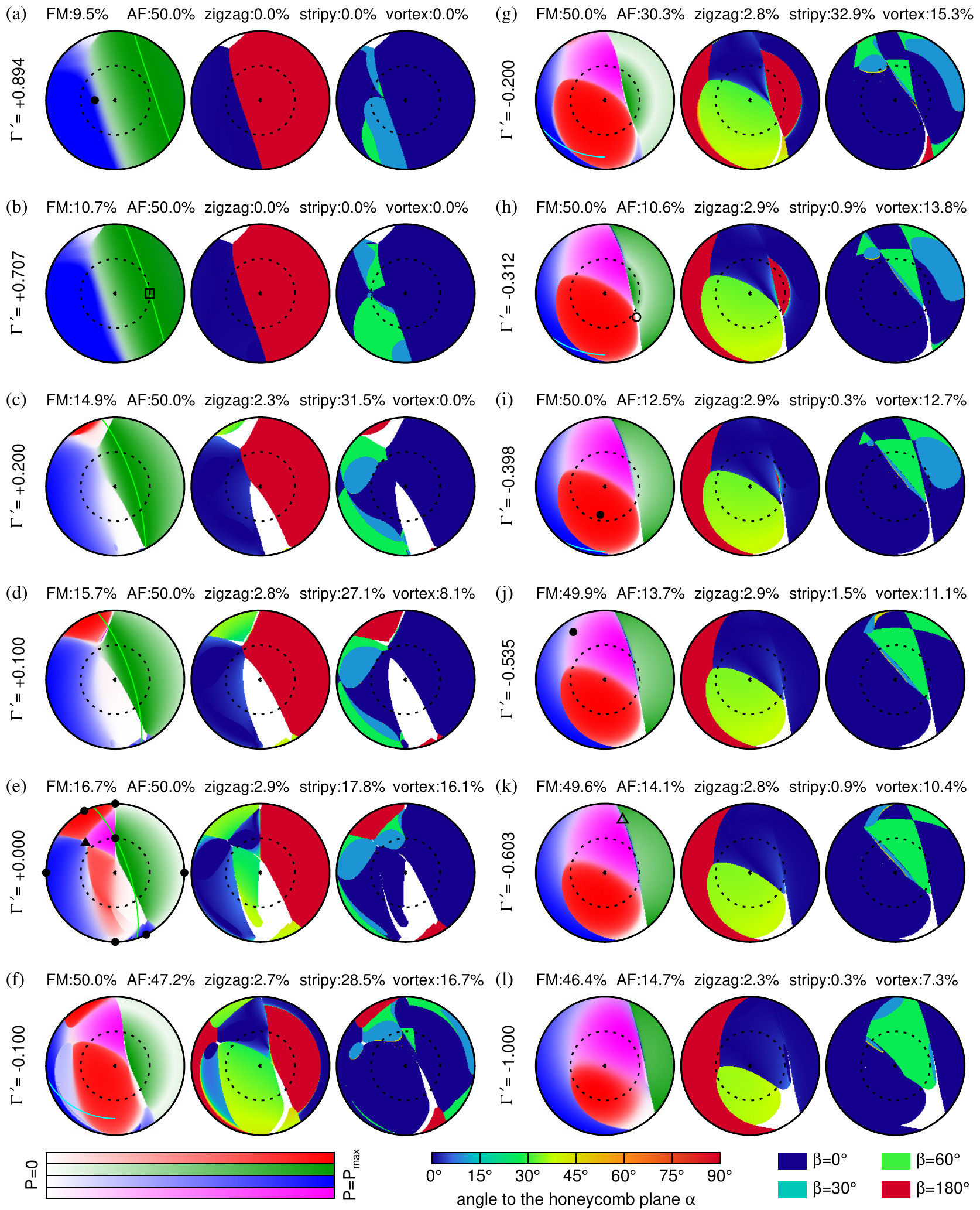}
\caption{
Phase diagram slices for selected $\Gamma'$ values and $J_3=0$. Left
circle in each panel shows a colormap of the probabilities of
the patterns: FM (blue), AF (green), zigzag (red),
stripy (blue), vortex (violet). The ranges of the corresponding scales (bottom
left) are determined by the maximum probability $P_\mathrm{max}$
as given at the top of the individual panels. Middle circles 
show the angle to the honeycomb plane (common scale at bottom center).
Right circles show the azimuthal angle $\varphi$ within the honeycomb plane.
Fluctuation-free lines given by Eq.~\eqref{eq:ffcondFM} (FM phase, light
blue) and \eqref{eq:ffcondAF} (AF phase, light green)
are indicated along with the special points: (hidden) SU(2) symmetry points
[$\bullet$ in panels (a),(e),(i),(j)], Kitaev points [$\bullet$ in (e),
$\circ$ in (h)], Ising point [$\Box$ in (b)], compass
point [$\blacktriangle$ in (e)] and compass-like point [$\vartriangle$
in (k)].
}
\label{fig:parevol}
\end{figure*}

\vspace{-2mm}\section{Ising-Kitaev-compass model -- derivation}\vspace{-2mm}
\label{sec:appendixIKc}

The EKH Hamiltonian has a single matrix form for all bond directions when
written in the rotating reference frame
$\tilde{x}_\gamma,\tilde{y}_\gamma,\tilde{z}_\gamma$:
\begin{equation}\label{eq:extKHtilde}
  \widetilde{H}=
  \begin{pmatrix}
    J-\Gamma & 0 & 0 \\[0.8em]
    0 & J+\tfrac{2K}{3}+\tfrac{\Gamma}{3}-\tfrac{4\Gamma^\prime}{3} 
        & \tfrac{\sqrt{2}}{3}(K-\Gamma+\Gamma^\prime) \\[0.8em] 
    0 & \tfrac{\sqrt{2}}{3}(K-\Gamma+\Gamma^\prime) 
        & J+\tfrac{K}{3}+\tfrac{2\Gamma}{3}+\tfrac{4\Gamma^\prime}{3}
  \end{pmatrix}.
\end{equation}
The matrix consists of two blocks, therefore one principal axis is
$\tilde{x}_\gamma$ and the other two lie in the
$\tilde{y}_\gamma\tilde{z}_\gamma$ plane. The eigenvalues of this matrix are:
\begin{align}
  \lambda_1 &= J-\Gamma, \notag \\
  \lambda_2 &= J+\tfrac12( K+\Gamma+\sqrt{(K-\Gamma)^2+8\Gamma^{\prime 2}}), \notag \\
  \lambda_3 &= J+\tfrac12( K+\Gamma-\sqrt{(K-\Gamma)^2+8\Gamma^{\prime 2}}).
\end{align}
If only the first eigenvalue $\lambda_1=J=\Gamma$ is non-zero, we obtain the
120$^\circ$ honeycomb compass model, the situation discussed in Section
\ref{sec:kitaevisingpoint}. For $\Gamma>0$, one can find a~parameter region
where only $\lambda_2$ is non-zero:
\begin{equation}\label{eq:kitaevisingplus}
  J=\Gamma,\; J(J+K)=\Gamma^{\prime 2},\; J>0,\; J+K>0.
\end{equation}
Analogous to this case, for $\Gamma<0$ only $\lambda_3$ is non-zero in the
region given by
\begin{equation}\label{eq:kitaevisingminus}
  J=\Gamma,\; J(J+K)=\Gamma^{\prime 2},\; J<0,\; J+K<0.
\end{equation}
The sole non-zero eigenvalue in both of these cases equals $3J+K$ and its sign
is the same as the sign of $\Gamma$, hence in the case $\Gamma>0$ used in the
main text $3J+K>0$ and the interaction is antiferromagnetic. The parameter
region \eqref{eq:kitaevisingplus} [or \eqref{eq:kitaevisingminus}] can be
parametrized by a single variable, which we choose as the angle $\vartheta$
between the $\tilde{z}_\gamma=Z$ axis and the interaction direction. In terms
of the model parameters, $\vartheta$ is given by 
\begin{equation}\label{eq:tan2the}
  \tan 2\vartheta=2\sqrt{2}\left(\frac{9\Gamma'}{J-K+8\Gamma'}-1\right) .
\end{equation}

It is also easy to obtain the parameters $J$, $K$, $\Gamma$, $\Gamma'$ 
realizing the model \eqref{eq:kitaevising_gen} with arbitrary $\vartheta$ and
$\varphi=\pi/2$:
\begin{align}
  J=\Gamma&=\tilde{K}\,\tfrac16\bigl[1+\cos\vartheta(\cos\vartheta-\sqrt{8}\sin\vartheta)\bigr],\\ 
  K       &=\tilde{K}\,\tfrac12\bigl[1-\cos\vartheta(\cos\vartheta-\sqrt{8}\sin\vartheta)\bigr],\\
  \Gamma' &=\tilde{K}\,\tfrac13\bigl[\cos2\vartheta+\tfrac1{\sqrt{8}}\sin2\vartheta)\bigr].
\end{align}

Finally, as demonstrated in Ref.~\cite{Cha15}, the nearest-neighbor EKH
model preserves its form under global rotation of the spin axes by $\pi$
around the $(111)$ direction. This transformation, labeled as $\mathcal{T}_1$
in Ref.~\cite{Cha15}, maps the model parameters $JK\Gamma\Gamma'$ onto
another set according to Eq.~(4) of Ref.~\cite{Cha15}. In the context of
the Ising-Kitaev-compass model, the rotation by $\pi$ around $(111)$ axis
corresponds to a sign change $\vartheta\rightarrow -\vartheta$. Indeed,
inserting the transformed set of parameters into Eq.~\ref{eq:tan2the}, one
observes the sign change. Obviously, the $\mathcal{T}_1$ transformation does
not have any effect in the Ising ($\vartheta=0$) and compass/``compass''
($\vartheta=\pm\pi/2$) case. Apart from these cases, it connects various pairs
of points on the curves in Fig.~\ref{fig:kitaevising}(c)-(e), most
importantly, the Kitaev model and its $\pi$-rotated variant.


\vspace{-2mm}\section{Detailed evolution of the phase diagram for nonzero $\Gamma'$}\vspace{-2mm}
\label{sec:appendixPD}

Figure~\ref{fig:parevol} presents a detailed evolution of the phases of the
nearest neighbor model ($J_3=0$) with the parameter $\Gamma'$ attaining both
positive and negative values. It was obtained using the method of
Sec.~\ref{sec:methcoh} for the hexagon-shaped 24-site cluster. The $\Gamma'$
values were selected to include all the special symmetry points discussed in
Ref.~\cite{Cha15} and in the present paper. The lines of
fluctuation-free ground states discussed in Sec.~\ref{sec:T6} are also
indicated.

\begin{figure}[b]
\includegraphics[scale=1.0]{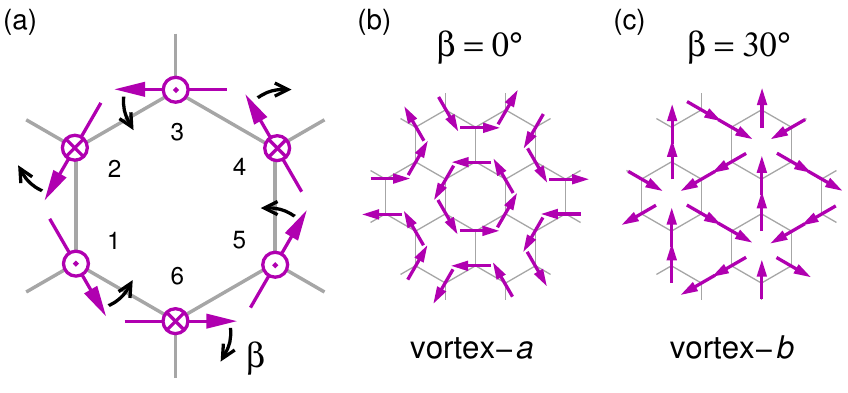}
\caption{
(a) Moment directions defined by Eq.~\eqref{eq:vortexAns} capturing vortex
in-plane structure and out-of-plane AF staggering. For the in-plane deviation
angle $\beta=0^\circ$ and $\beta=30^\circ$, we get the patterns (b)
\mbox{vortex-\textit{a}} and (c) \mbox{vortex-\textit{b}} of
Fig.~\ref{fig:prob}(b), respectively.
}
\label{fig:vortpattern}
\end{figure}

The plots use the same parametrization of the interactions as that of
Fig.~\ref{fig:prob} and show also the moment direction in the form of an
angle to the honeycomb plane $\alpha$ and an azimuthal angle in the honeycomb
plane $\beta$. Using the $XYZ$ reference frame of Fig.~\ref{fig:T6}(b), the moment
direction in the collinear phases is given by
\begin{equation}
  \vc n = (\cos\alpha\cos\beta,\cos\alpha\sin\beta,\sin\alpha).
\end{equation}
For the moments lying in the honeycomb plane, $\beta=0^\circ$ corresponds to
the direction perpendicular to a bond while $\beta=30^\circ$ is in-bond
direction. The values $\beta=60^\circ$ and $\beta=180^\circ$ are used in
the out-of-plane cases where a further distinction is necessary. In the vortex
phase, the directions of the moments at the six sublattices marked in
Fig.~\ref{fig:T6}(c) are
\begin{equation}\label{eq:vortexAns}
 \vc n_k = \bigl(\,
  \cos\alpha\cos\beta_k, \cos\alpha\sin\beta_k, (-1)^{k-1}\sin\alpha \,\bigr),
\end{equation}
where $\beta_k=(-1)^{k-1}\beta-60^\circ k$ and $k=1,2,\ldots 6$ labels the
sublattice. This Ansatz, depicted in Fig.~\ref{fig:vortpattern}, captures 
both patterns \mbox{vortex-\textit{a}}
($\beta=0^\circ$ or $\beta=60^\circ$) and \mbox{vortex-\textit{b}}
($\beta=30^\circ$) shown in Fig.~\ref{fig:prob}(b) and AF staggering in the
direction perpendicular to the honeycomb plane (nonzero~$\alpha$).

Not indicated in Fig.~\ref{fig:parevol} are the regions of the instability of
the zigzag phase observed in Fig.~\ref{fig:prob}(a) and
Fig.~\ref{fig:angles}(a),(b). Apart from the ones shown earlier, the regions
with non-collinear tendencies appear for large negative $\Gamma'\lesssim -0.5$
near the meeting point of the zigzag phase with FM and vortex phases and, to a
smaller extent, also at the bottom near the white region in
Fig.~\ref{fig:parevol}(j)-(l). Since the smaller zigzag phase visible in
Fig.~\ref{fig:parevol}(c)-(f) is a copy of the larger zigzag phase linked by
the exact $\mathcal{T}_1$ transformation, it also contains such patches of
instability. Similarly, the top and bottom white areas hosting incommensurate
orderings [visible in Fig.~\ref{fig:parevol}(a)-(c) and
Fig.~\ref{fig:parevol}(c)-(f), respectively] are linked by the
$\mathcal{T}_1$ transformation. The white area of
Fig.~\ref{fig:parevol}(j)-(l) maps to the $\Gamma<0$ case.


\end{document}